\documentclass[conference]{IEEEtran}

\def\showcopyright{1}
\def\showanalysis{1}

\usepackage[english]{babel}

\usepackage{url}

\usepackage{graphicx}
\usepackage{diagbox}

\usepackage{xspace}
\usepackage{xcolor} % Used to highlight comments
\usepackage{tabularx}
\usepackage{subfig}
\usepackage{enumitem}
\setlist{nosep}
\usepackage[T1]{fontenc}
\usepackage{hyperref}
\usepackage{amsmath}
\usepackage{amsthm}
\usepackage{amsfonts}

\usepackage{mathtools}
\usepackage{tikz}
\usepackage{soul}
\usepackage{multirow}
\usepackage{booktabs}
\usepackage{csquotes} % Used for easy quotation marks
\usepackage[normalem]{ulem} % For strike-through text
\usepackage[capitalize,noabbrev]{cleveref}

\crefname{lstlisting}{Listing}{Listings}
\Crefname{lstlisting}{Listing}{Listings}

\usepackage{listings}
\usepackage{siunitx} 
\usepackage{pdflscape}

\usepackage{etoolbox}
\preto\subequations{\ifhmode\unskip\fi}

\usepackage{sourcecodepro}
\usepackage{algorithm}
\usepackage{algpseudocode}
\newcommand{\name}{\textsc{G-SINC}\xspace}

\newcommand{\ts}{TS}

\usepackage[symbol]{footmisc}

\newtheorem{theorem}{Theorem}[section]

\newtheorem{lemma}[theorem]{Lemma}
\newtheorem{corollary}[theorem]{Corollary}

\renewcommand{\paragraph}[1]{\noindent\textbf{#1.}}

\newcounter{lineno}

\newcommand{\SSS}{\mathcal{S}}

\PassOptionsToPackage{hyphens}{url}

\usepackage[breakable]{tcolorbox}
\tcbuselibrary{breakable}
\definecolor{lightgray}{gray}{0.92}
\definecolor{darkgray}{gray}{0.50}

\newcommand{\greybox}[1]{%
	\begin{tcolorbox}[breakable, colback=lightgray, colframe=darkgray, left=1pt,right=1pt,top=0pt,bottom=0pt]%
		#1%
	\end{tcolorbox}%
}

\newcommand{\greyborder}[1]{%
	\begin{tcolorbox}[breakable, colback=white, colframe=darkgray, left=1pt,right=10pt,top=0pt,bottom=0pt]%
		#1%
	\end{tcolorbox}%
}

\ifnum\showcopyright=1
\usepackage{fancyhdr}
\fancypagestyle{firstpage}{
    \fancyhf{}

    \fancyfoot[C]{\vspace{-8pt}\centering\footnotesize{\textcopyright\ 2022 IEEE. Personal use of this material is permitted. Permission from IEEE must be obtained for all other uses, in any current or future media, including reprinting/republishing this material for advertising or promotional purposes, creating new collective works, for resale or redistribution to servers or lists, or reuse of any copyrighted component of this work in other works.}}
}
\else
\fi

\begin{document}

\title{G-SINC: Global Synchronization Infrastructure for Network Clocks}

\author{
    \IEEEauthorblockN{Marc Frei\IEEEauthorrefmark{1}, Jonghoon Kwon\IEEEauthorrefmark{1}, Seyedali Tabaeiaghdaei\IEEEauthorrefmark{1}, Marc Wyss\IEEEauthorrefmark{1}, Christoph Lenzen\IEEEauthorrefmark{2}, and Adrian Perrig\IEEEauthorrefmark{1}}
    \\
    \IEEEauthorblockA{\IEEEauthorrefmark{1}Network Security Group, Department of Computer Science, ETH Zurich, Switzerland
    \\Email: \{marc.frei, jong.kwon, seyedali.tabaeiaghdaei, marc.wyss, adrian.perrig\}@inf.ethz.ch}
    \IEEEauthorblockA{\IEEEauthorrefmark{2}CISPA Helmholtz Center for Information Security, Germany
    \\Email: lenzen@cispa.de}
}

\maketitle

\ifnum\showcopyright=1
\thispagestyle{firstpage}
\else
\fi

\begin{abstract}
Many critical computing applications rely on secure and dependable time which is reliably synchronized across large distributed systems. Today's time synchronization architectures are commonly based on global navigation satellite systems at the considerable risk of being exposed to outages, malfunction, or attacks against availability and accuracy. This paper describes a practical instantiation of a new global, Byzantine fault-tolerant clock synchronization approach that does not place trust in any single entity and is able to tolerate a fraction of faulty entities while still maintaining synchronization on a global scale among otherwise sovereign network topologies. Leveraging strong resilience and security properties provided by the path-aware SCION networking architecture, the presented design can be implemented as a backward compatible active standby solution for existing time synchronization deployments. Through extensive evaluation, we demonstrate that over 94\% of time servers reliably minimize the offset of their local clocks to real-time in the presence of up to 20\% malicious nodes, and all time servers remain synchronized with a skew of only 2 ms even after one year of reference clock outage.


\end{abstract}

\section{Introduction}
\label{sec:intro2}
Secure and dependable time synchronization is an essential prerequisite for many industries with applications in finance, telecommunication, electric power production and distribution, or environmental monitoring.

Current best practice to achieve large-scale time synchronization relies on Coordinated Universal Time (UTC) as a global time standard which is distributed within multiple, hierarchical synchronization networks from a set of primary time servers to every end host in the system. The Network Time Protocol (NTP)~\cite{rfc-5905} is a commonly used protocol for this purpose. Global Navigation Satellite Systems (GNSSes) are the most practical and cost effective sources of UTC as a reference time in this architecture.

Alternatives to GNSSes as reference clocks for UTC exist, but these more specialized methods are primarily used in purpose-built solutions with very specific use-cases:
\begin{itemize}
  \item Multi-Source Common-View Disciplined Clock (MSCVDC) is a recent design to support critical infrastructure systems that require fault-tolerant clock synchronization. In its current stage of development, MSCVDC makes use of GNSSes (and a central cloud service). Conceptually it would be feasible to augment the implementation with additional time sources, possibly even based on terrestrial transmitters. But it remains to be explored in how far the strict dependency on GNSSes could be removed for deployments beyond geographically restricted areas~\cite{commonviewgps2022, Lombardi2021}.
  \item Two-Way Satellite Time and Frequency Transfer (TWSTFT) provides significantly better synchronization quality than one-way communication with GNSSes at much higher complexity and cost~\cite{Sherman2021}. Like GNSSes, TWSTFT relies on satellite infrastructure which represents a potential source of disruption. The system also poses considerable unsolved scalability challenges to support more than a few tens of clients~\cite{twstft2022}.
  \item Time distribution over dedicated fiber optic links does not come with a dependency on satellites. At the same time this technology is not flexible enough to synchronize a potentially large and evolving set of primary time servers around the globe~\cite{Donley2020, Husmann2021}.
  \item Physical transport of high-accuracy clocks can be a viable synchronization method but it is even more limited in its application than dedicated fiber optic links~\cite{HellwigWainwright1975}.
\end{itemize}

We also note that installation of atomic clocks at every site to be synchronized is not sufficient to avoid dependencies on GNSSes. Atomic clocks can only serve as accurate and highly stable oscillators. As so-called frequency standards they cannot provide UTC by themselves but have to be continuously synchronized with external time sources~\cite{Lombardi12b}. Ongoing synchronization is required to compensate for nonidentical frequency drifts among separate clocks. Moreover, due to occasional (positive or negative) leap second insertions, UTC is not a fixed time scale that can be computed locally without external coordination.

Based on this short survey we conclude that there currently exists no practical alternative to GNSSes to receive UTC as a global reference time. In the context of critical infrastructure this means a significant risk since GNSSes exhibit a number of potentially severe vulnerabilities such as equipment failure~\cite{hubert2019}, misconfiguration, malicious manipulation by attackers~\cite{Nighswander2012}, and due to more localized spoofing or jamming attacks~\cite{Tippenhauer2011,hubert2020}. Natural disasters like, for example, solar superstorms could also hit and impact GNSSes~\cite{superstorm2021}.

Complete reliance on GNSSes in the traditional NTP architecture can be all the more serious in its consequences because primary time servers influenced by malfunctioning GNSSes will in turn affect the entire synchronization topology, and no fallback plan exists. It is apparent that time synchronization solely based on GNSSes does not fulfill fundamental dependability requirements for systems that serve indispensable functionalities in our society.

Related to these concerns, in 2020 NIST received a mandate to investigate possible approaches to the \textsl{deliberate, risk-informed use of positioning, navigation, and timing services} with the goal of supporting the needs of critical infrastructure owners and operators in the public and private sectors~\cite{govinfo2020}. The resulting report~\cite{Sherman2021} indicates an increasing awareness for this problem and highlights the importance of resilient time distribution. Manufacturers of time synchronization equipment are encouraged to further explore solutions in this space. In addition to the NIST work, there are a number of other publications that call attention to the high degree of dependence on GNSSes and the resulting economic impact in different industries~\cite{Lombardi2012,naspi2017,le2017impact,govuk2018gnss,RTI2019,enisa2020dependency,netnod2021why,Lombardi2021}.

Aiming to address the issues raised above, this paper describes \name: a novel, Byzantine fault-tolerant clock synchronization approach as a fully backward compatible extension of current time synchronization architectures. \name is executed among primary time servers across previously separate synchronization hierarchies and allows these top-level servers to reliably detect inconsistencies between time measurements retrieved via GNSSes and the globally synchronized time maintained by \name. In addition, the globally synchronized timing information can be used along the GNSS reference time as a redundant external time source for local clock corrections with the following qualities:
\begin{itemize}
  \item formally proven synchronization with real-time (UTC) for non-faulty nodes under normal conditions, with the accuracy of an arbitrary unreliable reference, e.g., a GNSS; this includes full recovery after transient faults of references;
  \item formally proven upper bound on the clock offsets among non-faulty nodes even under extreme conditions, where GNSSes are unavailable or cannot be trusted.
\end{itemize}

Our approach thus constructs a two-tier structure, where the top-level time servers run the global \name algorithm and all other intermediate time servers as well as end hosts then synchronize with their respective upstream providers as in currently deployed time synchronization hierarchies.

An additional contribution of this paper is the application of secure multipath communication in path-aware networking architectures~\cite{Yang2007, Godfrey2009, nebula2013, Castro2015, scion_cacm_2017} to improve fault tolerance and defend against on-path adversaries. It thereby helps to further advance methods to prevent time shifting attacks, orthogonal to approaches like Chronos~\cite{deutsch2018chronos, ietf-ntp-chronos-05}. We demonstrate this in an extensive evaluation based on a simulator framework that supports multipath communication at Internet scale. Our experiments show that over 94\% of time servers reliably minimize the offset of their local clocks to real-time even under the presence of 20\% malicious nodes, and that all time servers remain internally synchronized with a skew of at most 2 ms after one year of reference clock outage.

To implement the system, we make use of the SCION next-generation Internet architecture~\cite{scion_cacm_2017, SCIONBookv2}, which provides several mechanisms to realize the desired system properties. Besides the possibility to use multiple distinct paths in parallel, we highlight the fact that SCION paths are reversible and therefore symmetric. Hence, they help to increase time synchronization precision compared to clock offset measurements over the often asymmetric paths in today's Internet.

\section{Background}
\label{sec:background}
\subsection{Time Synchronization}
\label{ssec:timesync}
The goal of time synchronization is to limit the relative offsets among clocks to an acceptable range.
Due to the intrinsic error of clocks, computing devices usually cannot rely on an agreed upon clock configuration at a single point in time, but instead need to correct their clocks in a periodic manner.

Time synchronization algorithms can be divided into two categories: In \emph{external} algorithms, nodes synchronize with an independent outside clock source, while in \emph{internal} algorithms, nodes synchronize among themselves to an internally generated common time, e.g., to a leader or an averaged time.

Even though GNSS-based time synchronization provides high accuracy, it is not affordable to every system due to its relatively high cost, operational complexity, and energy consumption.
As an alternative, nodes can use the network infrastructure to synchronize their clocks to a reference server.
The most widely used clock synchronization protocol is NTP~\cite{rfc-5905}.
It approximates the relative clock offset between an NTP client and an NTP server by exchanging timestamps over the network.
The underlying assumption is that the round-trip delay divided by two is close to the one-way-delay in each direction.
Accurate time synchronization can therefore only be achieved when the one-way-delay of the packet's forward and backward path do not differ much, where the synchronization error amounts to half the difference between the forward and backward travel times.

In today's Internet, routing asymmetry, i.e., the phenomenon that paths in the forward and backward direction are different, is a widely observed phenomenon~\cite{he2005routing, de2015asymmetric, pathak2008measurement}. 
But even on a symmetric path, messages can encounter different buffer times~\cite{exel2014mitigation, levesque2015improving}, and random wire delay as well as software timestamping inaccuracies are another source of latency asymmetry~\cite{kannan2019precise}.
While NTP is simple and easy to adopt, its accuracy in large-scale deployments is therefore relatively low, meaning that it is limited to synchronization on the order of milliseconds even under ideal conditions~\cite{murta2006qrpp1}. NTP's limiting factors (with network asymmetry as a dominant source) are discussed further in~\cite{NovickLombardi2015}.

\subsection{Network Stability and Security}
\label{ssec:scion}

The correct functioning of network-based clock synchronization algorithms depends on the properties of the network architecture.
In case of the public Internet, the Border Gateway Protocol (BGP) is responsible for exchanging routing information among autonomous systems (ASes).
However, considering BGP's slow convergence after network failures~\cite{Labovitz2000, holterbach2019blink} as well as its lack of path stability and security, reliable and precise global clock synchronization is elusive in the current BGP-based Internet.
From those shortcomings, the next-generation architecture SCION~\cite{scion_cacm_2017, SCIONBookv2} emerges.

As a clean-slate network architecture, SCION has been designed with a focus on reliability and security. In contrast to BGP, network failures or misconfigurations cannot result in global outages, route hijacking is prevented by design, and network-based DDoS attacks are efficiently mitigated. Furthermore, SCION allows for trust heterogeneity and supports multipath communication for end hosts.

A fundamental building block of SCION is the concept of Isolation Domains (ISDs). An ISD constitutes a logical grouping of ASes with a uniform trust environment or a common jurisdiction.
Each ISD defines its own roots of trust and policies through a trust root configuration (TRC), combining a signed collection of certificates and policy specifications.
Within an ISD, the routing process is isolated from external attacks and misconfigurations.
A subset of ASes in an ISD, called core ASes, maintain connections to other ISDs.
Path exploration inside an ISD is separate from the routing process between ISDs; an end-to-end path is hence assembled by the source through (up to) three path segments: an up-segment from a regular AS to a core AS, a core-segment between core ASes, and a down-segment from a core AS to a regular AS.

Besides the multitude of apparent advantages, we also designed our system on top of SCION due to its real-world deployment~\cite{anapaya,cyrill2021deployment,sunrise,swisscom,switch}, its open-source implementation~\cite{scionproto}, and its global research testbed~\cite{Kwon2020}.

\subsection{Byzantine Fault Tolerance}
\label{ssec:byzantine}

Achieving global clock synchronization is a difficult task notably also because of the questionable trustworthiness of different actors.
Some synchronization peers may have malicious intentions and for example report wrong time information.
But even benign peers can accidentally misconfigure their infrastructure, suffer unexpected faults, or get compromised.

The Lynch-Welch algorithm~\cite{welch88} allows initially synchronized nodes in a fully connected network to avoid drifting apart and thus to keep closely synchronized local times, while being able to tolerate up to just under one third Byzantine node faults.
With $n$ participating nodes, the algorithm can therefore sustain $f$ Byzantine node faults, as long as $n \geq 3f + 1$.
Byzantine faults refer to the most general type of faults, meaning that a Byzantine node can be in an arbitrary state and send arbitrary messages at any time.
The algorithm executes in iterative rounds, where at the beginning of each round the nodes broadcast a message to all other nodes.
Every node then waits for a certain time period that depends on the maximum message transmission delay, in order to collect the arrival times of all messages from its synchronization peers, which it stores as a sorted array~$A$.
Each node then applies a fault-tolerant midpoint calculation, whose output is subsequently used to correct its local clock.
The function discards the $f$ smallest and the $f$ largest values of~$A$ and computes the arithmetic mean of the minimum and maximum of the remaining values.
This whole procedure is then repeated after a certain time $P$, which marks the beginning of the next synchronization round.

\section{Design Principles}
\label{sec:design}
\subsection{Goals and Challenges}
\label{ssec:goal}

The fundamental goal of this work is for collaborative networks without complete mutual trust to introduce a reliable time synchronization infrastructure.
However, our intention is not to introduce a new time synchronization protocol. Various such protocols achieving high accuracy have already been introduced. The main obstacle for them to achieve global time synchronization is the hierarchical synchronization structure stemming from a single global root of trust: a lack of resilience to abnormal input from the root. 
To overcome this, we seek the following architectural design:
(1) multi-source time synchronization rather than a single trust root;
(2) a distributed time synchronization structure that scales to the Internet;
and (3) a flexible architecture where existing time synchronization protocols can be leveraged.
This design raises the following research challenges.

\paragraph{Fault Tolerance} 
Multi-source time synchronization decreases the structural weakness of single root-driven time synchronization. That is, each node collects and analyzes various inputs from multiple time sources, calculates a consolidated clock offset, and performs local clock correction. This aims to guarantee resilience against arbitrary faults or malicious behavior by a subset of entities.

\paragraph{Load Distribution} 
In the context of Internet-scale multi-source time synchronization it
is impractical for each node to synchronize with all other nodes. Therefore, for scalable multi-source time synchronization, (1) the network needs to be hierarchically layered according to the requirement of synchronization accuracy, and (2) network segmentation is needed to distribute the load required for the entire global time synchronization.

\paragraph{Backward Compatibility} 
In order for a new global time synchronization approach to be rapidly deployed and ready to use, architectural continuity with existing time synchronization systems is required. By utilizing already established time synchronization resources (e.g., GNSSes, servers, and protocols), architectural continuity with the existing system and the new goal, reliable global time synchronization, are simultaneously achieved.

\subsection{System Model}
\label{ssec:model}
\paragraph{Topology}
We model a distributed system as a graph $G = (V, E)$, where $V$ is the set of nodes, and $E$ is the set of bidirectional communication links. 
An unknown subset $F\subset V$ of the nodes is (Byzantine) faulty. Faulty nodes may violate the protocol in an arbitrary manner. In the following, denote by $V_g:=V\setminus F$ the subset of \emph{correct} nodes.
Each node $v$ can send network packets to its neighbors $N_v := \{u \in V | \{v,u\} \in E\}$. 
For $u \in N_v$, $v$ could have parallel links that can be distinguished by means of an interface identifier or a port number. 
A \textit{path} is a sequence of nodes and links which are all distinct, i.e., $p = v_0, e_1, ..., v_{k-1}, e_k, v_k$ where $e_i = \{v_{i-1}, v_{i}\}, 1 \leq i \leq k$. 
Non-adjacent nodes $v,w\in V$ need to rely on the internal nodes of one or more paths from $v$ to $w$ to communicate.

\paragraph{Latency}
Communication suffers from delay, which might vary due to link congestion, queuing delay, and process scheduling, ranging in $[d-u, d]$.
Precise bounds on the (maximum) \emph{delay} $d$ and the delay \emph{uncertainty} $u$ might be hard to determine and these parameters might be different for different links in practice.
However, note that we just ask that delays are between $d-u$ and $d$, so overestimating these parameters is acceptable.

\paragraph{Hardware Clocks}
To measure the progress of time, each node $v$ is equipped with a hardware clock, which we model by an increasing function $H_v\colon \mathbb{R}^+_0 \rightarrow \mathbb{R}^+_0$ mapping the real time $t$ to the \emph{local time $H_v(t)$} at $v$ at time $t$:
\begin{equation}
H_v(t) = H_v(0) + \int_0^t h_v(\tau)\;d\tau,
\end{equation}
where $H_v(0) \in \mathbb{R}^+_0$ is the \textit{hardware offset} and $h_v(\tau)$ is the clock rate of $v$'s hardware clock.
We stress that $v$ has no access to $t$; it can only measure the progress of time by querying and storing $H_v(t)$.
Thus, the hardware clock rate $h_v(\tau)$ during a time interval $[t,t']$ determines how far measuring $t'-t$ via evaluating $H_v(t')-H_v(t)$ is off the mark.
In general, $h_v$ may vary over time depending on environmental conditions such as the ambient temperature, the stability of the supply voltage, or crystal quartz aging.
We assume that the \emph{clock drift} relative to real time is bounded by $\vartheta-1\ll 1$, i.e.,\footnote{For notational convenience, we assume a one-sided error here. If $H_v$ satisfies $1-\rho \le h_v(\tau) \le 1+\rho$ for some $\rho\ll 1$ and all $\tau$, then $H_v'(t):=H_v(0)+\int_0^t h_v(\tau)/(1-\rho)\;d\tau$ meets our requirements for $\vartheta = (1+\rho)/(1-\rho)\approx 1+2\rho$.}
\begin{equation}
\forall{v} \in V~\forall{t}\in \mathbb{R}^+_0~: 1\le h_v(t) \leq \vartheta.
\end{equation}
A relatively cheap quartz oscillator clock exhibits a drift of less than \SI{20}{ppm}~\cite{tirado2019performance} which corresponds to around 1.7 seconds per day, while a higher-end rubidium clock may have a drift as small as 
\SI{8}{ms} in a year~\cite{meinberg2017oscillator}.

\paragraph{Synchronization Requirements}
A clock synchronization algorithm computes at each correct node $v\in V_g$ a logical clock $L_v := \mathbb{R}^+_0 \rightarrow \mathbb{R}^+_0$.
``Good'' logical clocks should behave as closely as possible to ideal clocks.\footnote{The term \enquote{logical clock} is used here as a common term in the clock synchronization literature. Introduced, e.g., in~\cite{Lamport1985}, it is defined as the value of a hardware clock plus some correction value.}
However, since it is impossible to track the real time $t$ perfectly, different features of such ideal clocks result in different, competing requirements:
\begin{itemize}
  \item \textbf{accuracy:} minimize $\mathcal{A}(t):=\max_{v\in V_g}\{|L_v(t)-t|\}$;
  \item \textbf{skew:} minimize $\mathcal{G}(t):=\max_{v,w\in V_g}\{L_v(t)-L_w(t)\}$; and
  \item \textbf{bounded rates:} minimize $\mu$ s.t.\ $\forall t\in \mathbb{R}^+_0, \forall v\in V_g\colon$\linebreak $(1-\mu)h_v(t)\le \ell_v(t)\le (1+\mu)h_v(t)$, where $\ell_v:=\frac{dL_v}{dt}$.
\end{itemize}

\paragraph{Reference Time}
To achieve bounded accuracy, the nodes need some access to a reference tracking real time with bounded error. While, both from the viewpoint of philosophy and physics, it is not clear how ``real'' time should be defined, in the context of our work we consider UTC to equal the real time $t$. The nodes learn about $t$ via oracle access to a clock reference, which could be implemented by receiving time information from one or multiple
GNSS services. Under regular conditions, the oracle function $r_v$ at $v$ evaluated at time $t$ satisfies that $|r_v(t)-t|\le \varepsilon$, where $\varepsilon>0$ can be expected to be much smaller than $u$. In this case, simply regularly querying the oracle and interpolating between the returned values yields a logical clock with small $\mu$, where accuracy and skew are in $O(\varepsilon)$, because $L_v(t)-L_w(t)\le |L_v(t)-t|+|L_w(t)-t|\le 2\mathcal{A}(t)$ for all $v,w\in V_g$ and $t$. However, there is no guarantee that $r_v$ behaves this way at all times.

\paragraph{Objective}
The task of the clock synchronization algorithm is to make a best effort in leveraging $r_v$, in that we want to achieve $\mathcal{A}(t)=O(\varepsilon)\ll u$ when the oracle is reliable, while $\mathcal{G}(t)$ remains bounded even if $r_v$ misbehaves arbitrarily. Moreover, we want $\mathcal{A}(t)$ to become small again after a temporary failure of the oracle. Note that, on the upside, the requirement of bounded rates ensures that a temporary failure of the oracle does not result in a large deviation of logical clocks from UTC. However, it also means that the duration to return to the correct time once the oracle functions correctly again must be proportional to the accumulated error as well. This is a deliberate design decision: applications that rely on a trusted time reference are likely to expect the reference time to not deviate significantly from its nominal rate.

\subsection{Threat Model}
\label{ssec:threatmodel}

We assume that the adversary can compromise fewer than one-third of primary time servers. 
The adversary has full control over its territory (i.e., compromised entities and corresponding links) and can eavesdrop, inject, intercept, delay, and alter the on-path packets with negligible latency inflation. 
Besides the presence of an active attacker, we also consider network failures due to, e.g., link congestion, misconfiguration, and physical errors that hamper reliable clock synchronization. A detailed discussion of the implications of the SCION multipath architecture in our threat model is presented in~\cref{ssec:sys-arch-path-aware-net}.

\section{Architecture Design}
\label{sec:system-architecture}
To achieve reliable time synchronization at global scale, we propose \name. This section describes architectural design details, including topology layout, group membership, and the novel multi-source synchronization algorithm.

\begin{figure}[t]
\begin{center}
\includegraphics[width=0.48\textwidth]{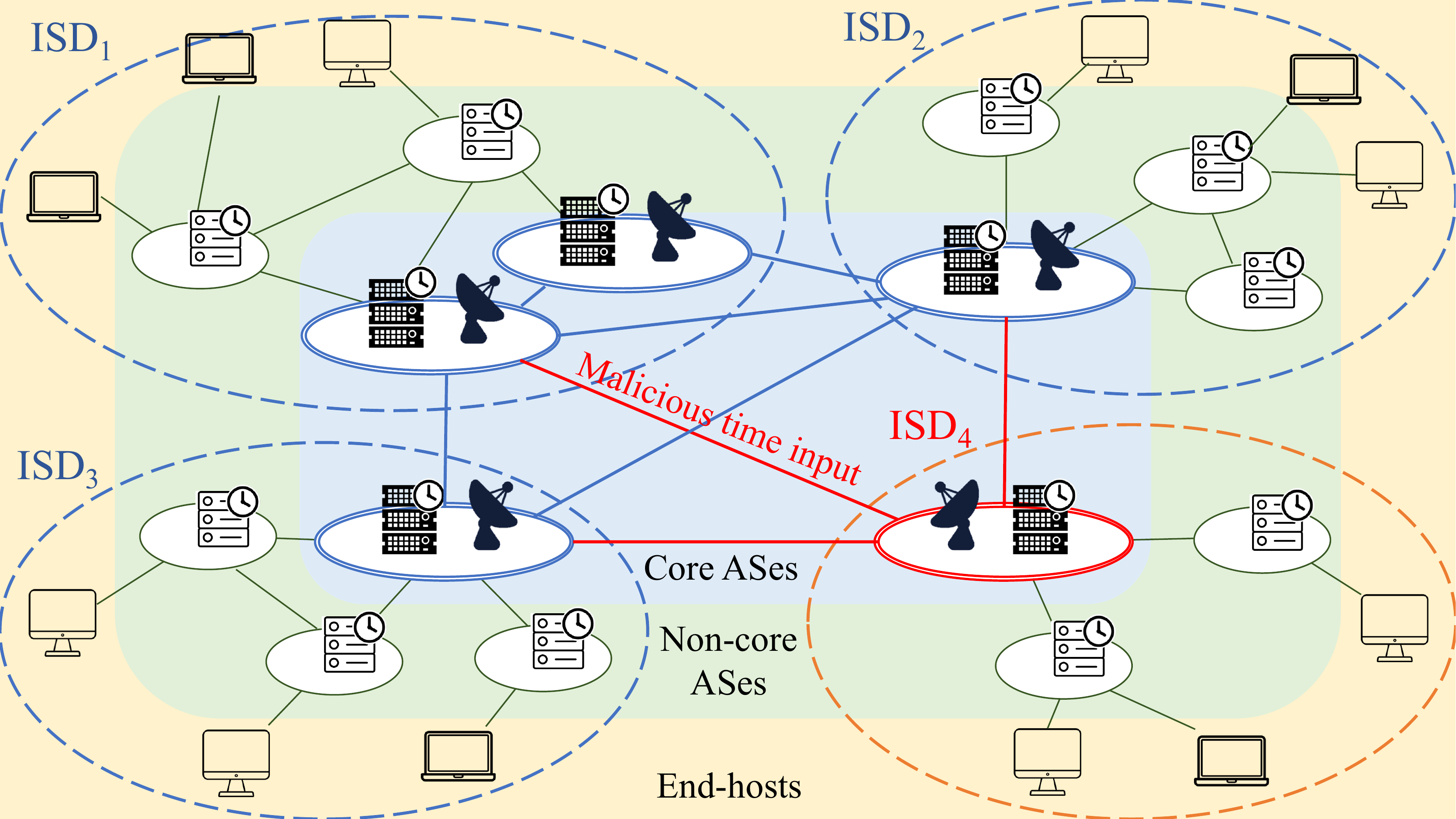}
\end{center}
\caption{\name architecture overview. Core ASes construct a core time synchronization network while each network cluster (ISD) operates in the traditional NTP structure.}
\label{fig:topology}
\end{figure}

\subsection{Architecture Overview}
\label{ssec:overview}

\name augments the traditional NTP architecture with the core concept of 
a Byzantine fault-tolerant algorithm to synchronize independent NTP networks. That is, 
\name ensures reliable and accurate time synchronization among primary time 
servers of each NTP network while the existing NTP network structure follows the established design. To this 
end, we summarize the key characteristics of our topology design as follows:

\begin{itemize}
  \item \emph{Hierarchical Partitioning}: We take advantage of the AS layering in SCION
  to form a two-tier structure. Core ASes run a global peer-to-peer 
  synchronization algorithm to achieve Byzantine fault tolerance. All other ASes 
  synchronize their time with the upstream providers to ensure scalability.
  \item \emph{Clustering}: The collaborative networks are clustered into ISDs based on their trust 
  relationship. The network clustering maintains the sovereign operation of ISDs, even 
  if external entities (e.g., GNSS providers) supply erroneous values, experience 
  outages, or attempt to interfere. An ISD can still maintain internal time 
  synchronization among its ASes and other ISDs operating correctly, as long as paths 
  through correctly operating ISDs exist.
  \item \emph{Logical N-to-N Peering}: Each core AS is virtually peered with all other 
  core ASes, securing multiple network-based time sources.
\end{itemize}

In \name, each core AS operates one or more primary time servers (\ts{es}). \cref{fig:topology} 
illustrates the basic topology. These core \ts{es}
also act as primary servers in 
the traditional NTP architecture and they typically use one or more GNSSes as their 
external time reference providing time values in UTC. 
We assume that core \ts{es} are equipped with local clocks driven by high-quality oscillators with an autonomous free run accuracy of $\pm$\SI{10}{\ms} (ca. \SI{0.3}{ppb}) or better 
after one year.

\subsection{Mutli-Source Time Synchronization}
\label{ssec:syncoverview}

\begin{figure}[t]
\begin{center}
\includegraphics[width=0.48\textwidth]{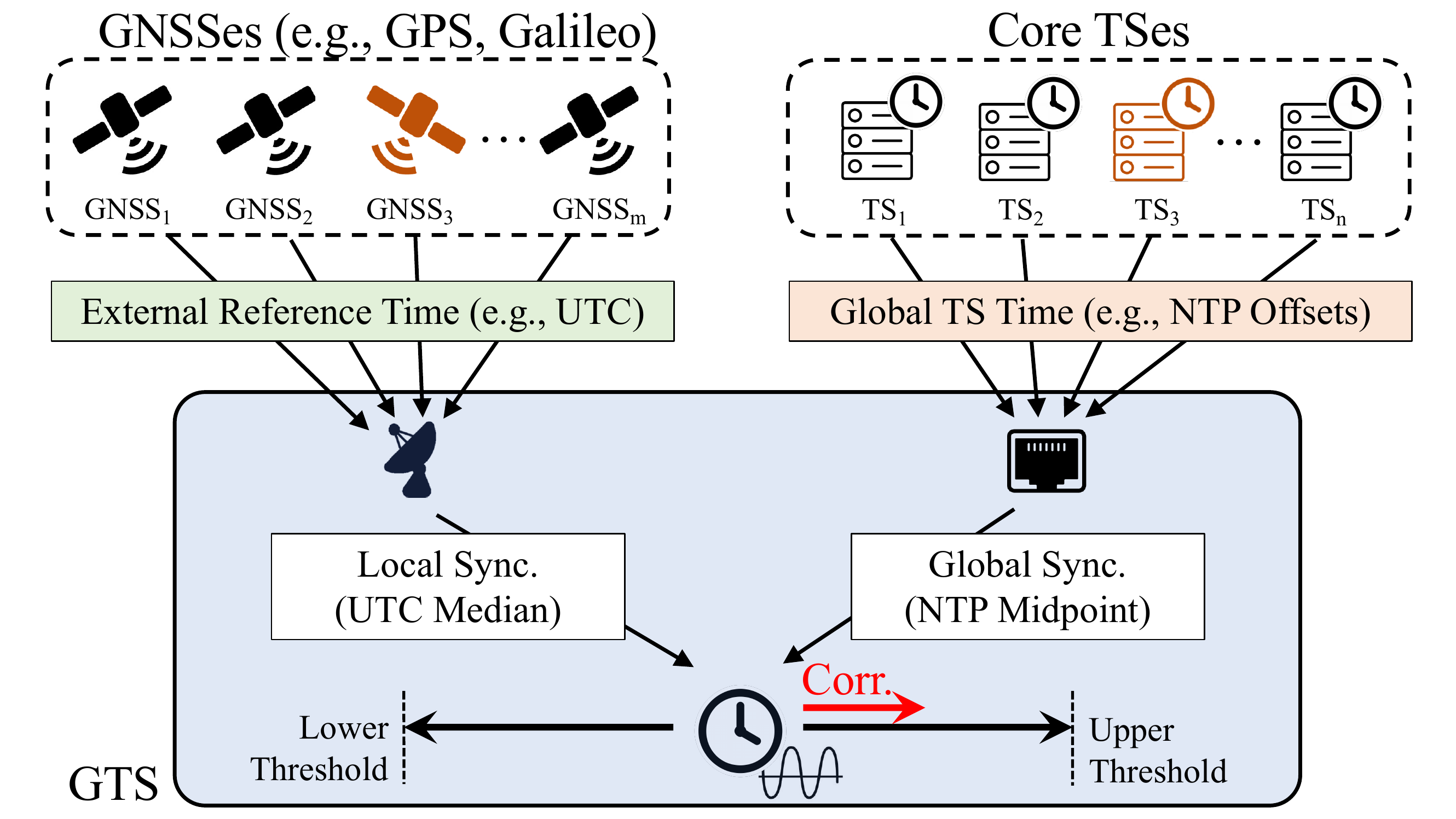}
\end{center}
\caption{A core TS synchronizes its local clock with two types of time sources: external reference times from GNSSes and NTP offsets to other core TSes.}
\label{fig:overview}
\end{figure}

The general strategy of our synchronization algorithm is to periodically compute 
an approximate agreement on clock correction values at each core \ts{} based 
on the relative clock offsets to its peers. Core \ts{es} then correct their local clocks 
towards the approximate global clock value. 
In addition to these periodic global synchronizations, each core \ts{} synchronizes the 
local clock with its local reference clocks. 
These local corrections are applied more frequently than the global clock corrections. By encapsulating 
the local synchronization in a separate process, we get an abstracted local clock that 
follows its reference clocks as long as these exhibit benign behavior. The abstraction also guarantees a maximum drift that is always within a constant factor of what 
the underlying hardware clock guarantees.
\Cref{fig:overview} illustrates this design.

The combination of global clock synchronization with the local synchronization process 
at each core \ts{} yields a globally synchronized time that is not only internally 
synchronized across all core ASes, but also externally synchronized to UTC as long as 
the UTC time values provided by the reference clocks are consistent with the globally 
synchronized time. If the offset between a local clock and the globally synchronized 
time exceeds a predefined threshold (on the order of the expected measurement error 
over the network) the \ts{} follows the globally agreed upon time to a larger degree 
than its reference clocks until the discrepancy is resolved.

Byzantine fault tolerance properties are achieved by deriving the algorithm from the 
extensively studied clock synchronization algorithm by Lynch and Welch~\cite{welch88}, 
see also \cref{ssec:byzantine}. The key idea is that in each round every participating 
node collects an array of relative clock offsets to each peer. Based on this data, a 
fault-tolerant midpoint (or averaging) function is computed, resulting in a global 
approximate agreement on the relative time differences among the core 
\ts{es}~\cite{dolev1986}. With this algorithm, the system is in theory able to tolerate 
faults or malicious behavior of up to one third of the nodes in the set of 
participating core \ts{es}.

\subsection{Consensus on Membership}
\label{ssec:membership}

A requirement for the core synchronization algorithm is that all core \ts
{es} agree on the same set of core \ts{es} to synchronize with. In SCION it is
possible to provide \ts{es} in the core network with a consistent enough view
of all core ASes in the network to satisfy this requirement in practice.
So-called trust root configurations (TRCs) are disseminated among all ISDs as
part of the core beaconing (a fundamental network functionality in SCION to
construct path segments among core ASes within an ISD and across ISDs). Each
TRC consists of a signed collection of cryptographic information entries,
including a list of the core ASes in a given ISD. The public-key
infrastructure (PKI) of SCION precisely defines policies, roles, and procedures
covering verification, update, and revocation of TRCs as well as recovery from
catastrophic events. The set of all core ASes can therefore be maintained based
on TRCs without introducing additional mechanisms besides what is already
provided by SCION's control-plane PKI. Since every core AS has to go through an
official approval process to be included in a TRC, we also substantially reduce
the risk of possible Sybil attacks~\cite{douceur2002} as a common threat in
peer-to-peer settings.

\algblock{Input}{EndInput}
\algnotext{EndInput}
\algblock{Algorithm}{EndAlgorithm}
\algnotext{EndAlgorithm}
\newcommand{\Desc}[2]{\State \makebox[2em][l]{#1}#2}

\renewcommand{\algorithmicrequire}{\textbf{Input:}}

\begin{algorithm}[t]
\caption{Local Clock Synchronization}\label{alg:local_sync}
\small
\begin{algorithmic}[1]
	\Input
		\Desc{$I$}{Interval between local clock synchronizations}
		\Desc{$X$}{Local sync. impact factor ($X$ > 1)}
	\EndInput
	\Algorithm
		\State \textit{maxCorr} $\gets X \cdot$ \textsc{LocalClock.MaxDrift}($I$)
		\State \textit{refTime} $\gets$ \textsc{ReferenceClock.Time}()
		\State \textsc{LocalClock.SetTime}(\textit{refTime})
		\While{true}
			\State \textit{refTime} $\gets$ \textsc{ReferenceClock.Time}()
			\State \textit{locTime} $\gets$ \textsc{LocalClock.Time}()
			\State \textit{loff} $\gets$ \textit{refTime} - \textit{locTime}
			\If{$|$\textit{loff}$|$ > 0}
				\State \textit{corr} $\gets$ sgn(\textit{loff}) $\cdot$ min($|$\textit{loff}$|$, \textit{maxCorr})
				\State \textsc{LocalClock.AdjustTime}(\textit{corr}, $I$)
			\EndIf
			\State \textsc{LocalClock.Sleep}($I$)
		\EndWhile
	\EndAlgorithm
\end{algorithmic}
\end{algorithm}

\begin{figure}[t]
\begin{center}
\includegraphics[width=0.42\textwidth]{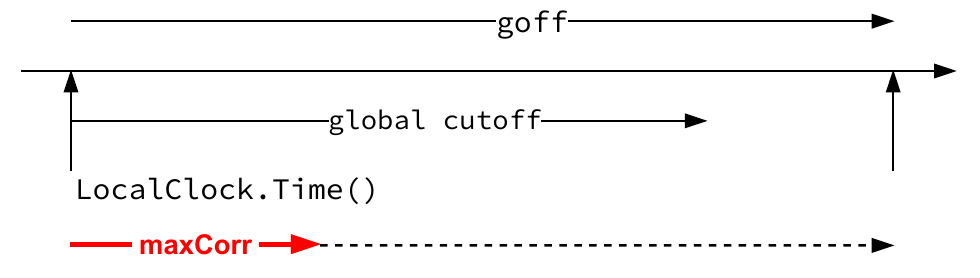}
\end{center}
\caption{Global Clock Correction}
\label{fig:global_sync}
\end{figure}

\subsection{Local Clock Synchronization}
\label{ssec:sys-arch-local-sync}

Each core TS internally runs a process that synchronizes the local clock with the connected external reference clocks, see~\cref{alg:local_sync}. The local clock and the external reference clocks are abstracted into separate modules: \textsl{LocalClock} and \textsl{ReferenceClock}. Every synchronization round begins by measuring the accumulated local offset \textsl{loff} as the difference between \textsl{ReferenceClock.Time()} and \textsl{LocalClock.Time()}. If $|$\textit{loff}$| > 0$, the corresponding correction value is computed and applied to \textsl{LocalClock}. However, this correction is limited to \textsl{maxCorr} in each round, so that a faulty or malicious reference cannot manipulate the local clock arbitrarily. We set \textsl{maxCorr} to the maximum drift the local clock may experience, scaled by a constant coefficient $X > 1$, to make sure that the reference clock is able to pull the local clock towards itself even if the local clock drifts maximally in the opposite direction. Corrections are thus capped by a function of the maximum expected time drift of the local hardware clocks. This important parameter is well-documented by the manufacturers of the intended clock sources and can also be independently tested. The local clock synchronization process repeats at interval $I$.

\begin{algorithm}[t]
	\caption{Global Clock Synchronization}\label{alg:global_sync}
  \small
	\begin{algorithmic}[1]
		\Input
		\Desc{$N$}{Number of nodes}
		\Desc{$F$}{Number of faulty nodes ($N >= 3F + 1$)}
		\Desc{$P$}{Set of synchronization peers}
		\Desc{$J$}{Interval between global clock}
		\Desc{}{synchronizations ($J >= I$)}
		\Desc{$G$}{Global cutoff}
		\Desc{$Y$}{Global sync. impact factor ($Y > X + 1$)}
		\EndInput
		\Algorithm
		\State \textit{maxCorr} $\gets Y \cdot$ \textsc{LocalClock.MaxDrift}($J$)
		\While{true}
			\State $M \gets$ [0] \Comment{Array of NTP offset measurements}
			\For{$p \in P$}
				\State $m \gets$ \textsc{PathAwareNTPOffset}($p$)
				\State $M \gets M$ + [$m$]
			\EndFor
			\State $M \gets$ sort($M$)
			\State \textit{goff} $\gets$ ($M$[$F$] + $M$[$N-1-F$]) / 2
			\If{$|$\textit{goff}$| > G$}
				\State \textit{corr} $\gets$ sgn(\textit{goff}) $\cdot$ min($|$\textit{goff}$|$, \textit{maxCorr})
				\State \textsc{LocalClock.AdjustTime}(\textit{corr}, $J$)
			\EndIf
			\State \textsc{LocalClock.Sleep}($J$)
		\EndWhile
		\EndAlgorithm
	\end{algorithmic}
\end{algorithm}

\subsection{Global Clock Synchronization}
\label{ssec:sys-arch-global-sync}
In parallel to the local synchronization process, a global synchronization algorithm is executed, which ensures that the clocks of all core 
\ts{es} stay internally synchronized.
Essential parameters of~\cref{alg:global_sync} are defined as follows: $N$ is the number of core \ts{} nodes 
participating in the synchronization and $F$ is the assumed maximum number of faulty 
nodes on which the Byzantine fault tolerance argument is grounded. Clock 
synchronization is performed in rounds. The constant $J$ specifies the time interval 
between global synchronizations.
As for the local synchronization, \textsl{maxCorr}
stores the maximum correction value that we are willing to 
apply within one synchronization round. This value has to be chosen large enough so 
that it can compensate for the entire offset that the local synchronization may 
introduce over the interval $J$. This amounts to the maximum expected clock drift over 
the interval $J$ plus the term $X \cdot LocalClock.MaxDrift(J)$ which is the maximum 
correction towards the reference clock that can accumulate on top of the local clock's 
intrinsic drift during the given interval $J$. Taken together 
this results in the scale factor $Y~(> X + 1)$ for the maximum clock drift.

The array \textsl{off} collects the relative clock offsets to the local \ts{} itself 
and to every peer \ts{} as measured by the the function \textsl{PathAwareNTPOffset}, 
see~\cref{alg:ntp_offset} and~\cref{ssec:sys-arch-path-aware-net}. Since the offset of 
any given local clock to itself is $0$, \textsl{off} is initialized with the value $0$. The relative clock offsets to the $N - 1$ peers are appended
subsequently. After sorting \textsl{off}, the approximate agreement
for the global clock offset  \textsl{goff} is computed by applying the Byzantine
fault-tolerant midpoint calculation proposed by Lynch and Welch~\cite{welch88}.

The resulting clock adjustment is controlled based on \textsl{goff}. If 
$|$\textit{goff}$|$ lies beyond the global cutoff value $G$, a clock correction towards
\textsl{goff} is computed so that the correction keeps the previously computed
 direction of \textsl{goff} but never  exceeds \textsl{maxCorr}, see~\cref
 {fig:global_sync}. Otherwise, we don't apply the global clock correction. $G$ is
 assigned a value on the order of the NTP measurement accuracy.

\textsl{LocalClock.Sleep} at the end of the loop suspends execution for the specified 
duration taking possible time adjustments issued previously via 
\textsl{LocalClock.AdjustTime} into account.

\paragraph{Formal Analysis}
\ifnum\showanalysis=1
An in-depth mathematical analysis of the synchronization routines is provided 
in~\cref{sec:formal-analysis}. As a key result, \cref{thm:worst_case_bound} states that even in 
a worst-case scenario where the reference clocks can behave arbitrarily, the \name{} 
clock skew is upper-bounded by $4\delta$ (plus negligible terms), where $\delta$ 
denotes the maximum expected offset measurement error over the network.
Moreover, \cref{thm:good_reference} guarantees that under normal conditions, the skew 
is bounded by the fourfold of the local clock's maximum drift during the synchronization interval~$I$.
\else
An in-depth mathematical analysis of the synchronization routines is provided 
in~\cite{g-sinc-extended-2022}. The first key result states that even in 
a worst-case scenario where the reference clocks can behave arbitrarily, the \name{} 
clock skew is upper-bounded by $4\delta$ (plus negligible terms), where $\delta$ 
denotes the maximum expected offset measurement error over the network.
A second theorem in the analysis guarantees that under normal conditions, the skew 
is bounded by the fourfold of the local clock's maximum drift during the synchronization interval~$I$.
\fi

\begin{algorithm}[t]
	\caption{Path-Aware NTP Offset Computation\\ (\textsc{PathAwareNTPOffset})}\label{alg:ntp_offset}
  \small
	\begin{algorithmic}[1]
		\Input
			\Desc{$p$}{Synchronization peer}
		\EndInput
		\Algorithm
			\State $D \gets$ \textsc{SCION.GetDisjointPaths}($p$)
			\State $M \gets$ [] \Comment{Array of NTP offset measurements}
			\For{$d \in D$}
				\State $t_0, t_1, t_2, t_3$, \textit{ok} $\gets$ \textsc{NTP.Measure}($p$, $d$)
				\If{\textit{ok}}
					\State $m \gets$ (($t_1 - t_0$) + ($t_2 - t_3$)) / 2
					\State $M \gets M$ + [$m$]
				\EndIf
			\EndFor
			\State \textit{off} $\gets 0$
			\If{$|M| > 0$}
				\State \textit{off} $\gets$ median(\textit{M})
			\EndIf
			\State return \textit{off}
		\EndAlgorithm
	\end{algorithmic}
\end{algorithm}

\subsection{Path-Aware Networking}
\label{ssec:sys-arch-path-aware-net}

A central concept of the SCION networking architecture is comprehensive path transparency and control that enables senders to simultaneously select multiple paths to carry packets towards the destination. In general, this multipath communication capability can be used to optimize bandwidth and latency as well as to enhance overall availability due to increased resilience against link failures, or to avoid untrusted infrastructure along the way. In the specific application of global time synchronization, multipath communication enables designing a system that is able to closely approximate optimal accuracy while also improving security and fault tolerance of the synchronization process even over the public Internet. 

\paragraph{Path Selection and Symmetry} When a sender is creating a packet to be sent over the network, it first queries a set of paths to the target end host. These paths are discovered and disseminated by the SCION control plane based on individual path segments at the level of ASes. At the end host path segments are then combined into actual end-to-end paths. For typical network topologies it is expected that the result set for a path query to a given target end host will consist of up to a few dozens of paths, especially between core ASes. The sender will thus select one or more paths out of the set of available paths, which can be used simultaneously even from endpoints connected by a single link. For each selected path, the sender creates packets that include path information in their headers as packet-carried forwarding state (PCFS)~\cite{scion_cacm_2017,path_awareness_2018}.
The destination host receiving those packets can either reply back to the source by fetching its own set of paths, or by reversing the PCFS information of the received packets.
The latter approach allows the response packets to traverse the same path but in the backwards direction.
This path symmetry can help reducing the latency variance between NTP requests and responses, leading to potentially higher measurement accuracy.

\paragraph{On-Path Adversaries}
The clock synchronization algorithm proposed by Lynch and Welch (\cref{ssec:byzantine}) allows to tolerate up to just under one-third arbitrary, i.e., Byzantine, faulty processes.
This holds under the assumption that the communication network is fully connected, and hence every process can communicate directly with all other processes.
This assumption is not met for the Internet however, where ASes are connected to others via many further (potentially malicious) ASes.
In order to account for this fact, we have to consider one of two endpoint ASes as malicious when the path between them is at least partially controlled by an attacker.

\paragraph{Threat Mitigations}
By leveraging a multipath-aware networking substrate, we strive to approximate a fully connected network by decreasing the impact that on-path adversaries can have on offset measurements.
First, paths via ASes considered untrustworthy can be filtered out and not be used for the NTP measurements.
Second, by only considering the median of NTP measurements conducted over multiple paths in~\cref{alg:ntp_offset}, many paths need to be compromised in order for an attacker to shift the resulting NTP offset.
To avoid letting a single malicious AS influence multiple paths, disjoint, i.e., non-overlapping, paths are preferred.
In some cases, choosing a single highly trusted path or a single path with the least number of ASes, or random selection might serve as an alternative strategy.
Given the topology dependence, we can only provide average-case numbers obtained though large-scale simulations to quantify the effect of faulty or adversarial network entities in the threat model. Our simulations show that, in realistic topologies, over 99\% of nodes in the core network maintain close synchronization with real-time in the presence of up to 10\% faulty nodes; with 20\% faulty nodes, over 94\% stay closely synchronized, see~\cref{ssec:reliability}. A thorough theoretical understanding of how general network topologies map to the best achievable resilience is an open research problem. Our approach allows to tackle this subproblem in a modular way as future work.

\paragraph{Synchronization Within a NTP Cluster}
With its multi-source synchronization algorithm among primary \ts{es} \name avoids complete reliance on GNSSes at the top-tier of the traditional NTP architecture. To achieve reliable end-to-end synchronization, intermediate (i.e., non-primary) \ts{es} and end-hosts also need to defend against faults and malicious actors on the path to upstream \ts{es}. In particular, the architecture also has to cope with malicious primary \ts{es} at the top of a hierarchy which could break synchronization of downstream nodes. We approach this part of the problem by combining the previously proposed Chronos mechanism~\cite{deutsch2018chronos, ietf-ntp-chronos-05} with path-aware networking concepts introduced in this section. Chronos is specifically designed to prevent time-shifting attacks in the presence of Byzantine faults in a NTP network. Enhanced with multipath support, this strategy is expected to result in significantly improved fault-tolerance compared to today's synchronization hierarchies based on unmodified NTP clients.

\paragraph{NTP Message Authentication}
To guarantee that only messages from verified members of the core synchronization group are processed, it is necessary to authenticate the request and response packets exchanged during the NTP-based offset measurements. NTP provides Network Time Security (NTS) as a cryptographic security mechanism for NTPv4 via extension fields in the NTP packet format and the NTS Key Establishment protocol (NTS-KE) to create and manage the corresponding key material between NTP clients and servers~\cite{rfc-8915}. It is entirely possible to use NTS over SCION and we successfully tested this in prototype deployments. The SCION architecture offers however a more lightweight and therefore preferred alternative in the form of DRKey~\cite{rot2020piskes}, which enables highly efficient authentication of all message exchanges on the network layer.

\ifnum\showanalysis=1
\section{Formal Analysis}
\label{sec:formal-analysis}
\allowdisplaybreaks

In the following, we present a formal analysis of the \name synchronization routines. We make the simplifying assumption that all nodes are taking part in the algorithm from the start and that the set of faulty nodes is static. As we discuss in \cref{ssec:membership}, core ASes can join (or leave) under nominal conditions, allowing the system to grow and adapt. Under worst-case conditions necessarily some limitations apply.

We will analyze the algorithm in terms of the offset of real times at which nodes start the $i$-th iteration of the loop in~\cref{alg:global_sync}.
Determining the most suitable interpolation method to obtain nicely behaved clocks that are synchronized at all times is beyond the scope of our formal analysis.

Denote by $L_v(t)$ the value that \textsl{LocalClock.Time()} would return if called at time $t$.
In order to specify the required behavior of \textsl{LocalClock.AdjustTime(offset: corr, duration: D)} being called at time $t_0$, consider a hypothetical scenario in which the call does not occur and denote the resulting clock by $L_v'(t)$.
For the sake of our analysis, we only require that
\begin{itemize}
  \item At times $t\ge t_0+D$, $L_v(t)=L_v'(t)+\mathrm{corr}$.
  \item At times $t\in (t_0,t_0+D)$, $L_v(t)\in [L_v'(t),L_v(t')+\mathrm{corr}]$.
  \item At times $t\le t_0$, $L_v(t)=L_v'(t)$.\footnote{This condition automatically holds due to causality; we merely list it for clarity and completeness.}
\end{itemize}
It is desirable that the derivative of $L_v(t)$, i.e., the instantaneous frequency, remains close to $1$ during $(t_0,t_0+D)$.
The benchmark here should be simple linear interpolation, i.e., evenly spreading the clock adjustment over the allocated period.
This avoids to unnecessarily increasing the skew the algorithm achieves during the adjustment process and increases the relative drift of the clock by no more than $\mathrm{corr}/D$.
However, in practice one might choose a different interpolation method to avoid a ``sudden'' transition in frequency or to achieve other desired properties.

Last, but not least, it should be stressed that $L_v(t)$ deviating from the above requirements is not problematic, so long as the deviation remains insignificant.\footnote{For example, the NTP clock discipline algorithm internally involves a control loop, which might slightly ``overshoot'' the target correction before swinging back.}
Any such deviation then simply adds (or, more precisely, might add) to the skews in actual runs compared to the outcome of our analysis.

%###
\subsection{Local Synchronization}
%###
Recall the local synchronization algorithm given in \cref{alg:local_sync}.
We first show that, in essence, this routine produces a new reference clock, which precisely tracks UTC so long as the reference does the same, while not drifting too much regardless of the behavior of the reference clock.
However, this statement must be qualified by the assumption that the global synchronization algorithm given in \cref{alg:global_sync} does not interfere.
It will be convenient for our analysis to express this as \cref{alg:local_sync} generating a local reference clock for use by \cref{alg:global_sync}.

To this end, denote by $R_v(t)$ the clock function obtained by pretending that no clock corrections are applied by \cref{alg:global_sync}.
Note that this yields a well-defined function, because we specified the behavior of \textsl{LocalClock.AdjustTime(offset: corr, duration: I)} relative to a fictional clock that is not adjusted.
\greyborder{
\begin{lemma}\label{lem:Rdrift}
For any $t_1\ge t_0$, it holds that
% \begin{equation*}
\begin{multline*}
|R_v(t_1)-R_v(t_0)-(t_1-t_0)|\le \\
 (\vartheta X+1)(\vartheta-1)(t_1-t_0)+X(\vartheta-1)I.
\end{multline*}
% \end{equation*}
\end{lemma}
}
\begin{proof}
Recall that the local clock satisfies $t'-t\le H_v(t')-H_v(t)\le \vartheta(t'-t)$ for all $t'\ge t$.
Hence, \textsl{LocalClock.MaxDrift(I)} returns $(\vartheta-1)I$.
The loop in \cref{alg:local_sync} ensures that always at most one correction is being applied (by this algorithm), which is applied over $I$ local time and has absolute value at most $X(\vartheta-1)I$.
Since $H_v(t_1)-H_v(t_0)\le \vartheta(t_1-t_0)$, at most $\lfloor \vartheta(t_1-t_0)/I\rfloor+1$ such corrections are applied within $t_1-t_0$ time.
Accordingly,
\begin{align*}
&\, |R_v(t_1)-R_v(t_0)-(t_1-t_0)|\\
\le &\, |H_v(t_1)-H_v(t_0)-(t_1-t_0)|\\
&\qquad+\left(\left\lfloor \frac{\vartheta(t_1-t_0)}{I}\right\rfloor+1\right)X(\vartheta-1)I\\
\le &\, (\vartheta-1)(t_1-t_0)+\vartheta (\vartheta-1)X(t_1-t_0)+X(\vartheta-1)I\\
= &\, (\vartheta X+1)(\vartheta-1)(t_1-t_0)+X(\vartheta-1)I.\qedhere
\end{align*}
\end{proof}
Thus, choosing $X$ small is beneficial in that the drift of $R_v$ is not going to be much larger than that of $H_v$.
In particular, up to an additive error that can be controlled by making $I$ small, for constant $X$ we ``increase'' the drift by no more than a factor of roughly $X+1$.

The reason for requiring that $X>1$ becomes apparent when analyzing the ability to track UTC when the reference accurately reflects it.\footnote{For simplicity, we assume that the reference exactly reproduces UTC here. An additional error of up to $\varepsilon>0$ simply adds to the difference. Typical GNSS receivers reproduce the ``true'' time with less than $100$\,ns error.}
\greyborder{
\begin{lemma}\label{lem:Roffset}
Suppose that at all times $t\in [t_0,t_1]$ calling \textsl{ReferenceClock.Time()} at node $v$ returns $t$.
Then for each such time $t$, it holds that
\begin{align*}
|R_v(t)-t|&\le \max\{|R_v(t_0)-t_0|+3X(\vartheta-1)I\\
&\qquad\;-(X-1)(\vartheta-1)(t-t_0),2(\vartheta-1)I\}.
\end{align*}
\end{lemma}
}
\begin{proof}
Let $t_0'\in [t_0,t_0+I]$ be the first time after $t_0$ when the loop in \cref{alg:local_sync} starts a new iteration;
since $H_v(t_0+I)-H_v(t_0)\ge I$, such a time exists.
For any $t\in [t_0,t_0']$, due to the limit on clock corrections we have that
\begin{align*}
&\,|R_v(t)-t|\\
\le&\,|R_v(t_0)-t_0|+X(\vartheta-1)I+|H_v(t)-H_v(t_0)-(t-t_0)|\\
\le&\, |R_v(t_0)-t_0|+(X+1)(\vartheta-1)I\\
\le&\, |R_v(t_0)-t_0|+2X(\vartheta-1)I-(X-1)(\vartheta-1)(t-t_0),
\end{align*}
showing the statement of the lemma for $t\in [t_0,t_0']$.

Denote for $i\in \mathbb{N}_{>0}$ by $t_i'$ the $i$-th time after $t_0'$ when a loop iteration completes.
Note that $t_i'-t_{i-1}'\le H_v(t_i')-H_v(t_{i-1}')=I$, so $t_i'-t_0'\le iI$.
We claim that
\begin{align*}
|R_v(t_i')-t_i'|&\le\max\{|R_v(t_0)-t_0|+2X(\vartheta-1)I\\
&\qquad\;-(X-1)(\vartheta-1)(t_i'-t_0),(\vartheta-1)I\}
\end{align*}
for all $t_i'\le t_1$, which we show by induction of $i$.
The base case is $i=0$, for which the claim follows from the already established bound for time $t_0'$.
For the step from $i-1$ to $i$, note that a clock correction of absolute value $\min\{|R_v(t_{i-1}')-t_{i-1}'|,X(\vartheta-1)I\}$ with the appropriate sign is applied.
Since $X>1$, we get that
\begin{align*}
&\,|R_v(t_i')-t_i'|\\
\le &\,\max\{|R_v(t_{i-1}')-t_{i-1}'|-X(\vartheta-1)I,0\}\\
&\qquad\;+|H_v(t_i')-H_v(t_{i-1}')-(t_i'-t_{i-1}')|\\
\le &\, \max\{|R_v(t_{i-1}')-t_{i-1}'|-X(\vartheta-1)I,0\}\\
&\qquad\;+(\vartheta-1)(t_i'-t_{i-1}')\\
\le &\, \max\{|R_v(t_{i-1}')-t_{i-1}'|-X(\vartheta-1)(t_i'-t_{i-1}'),0\}\\
&\qquad\;+(\vartheta-1)I\\
\le &\, \max\{|R_v(t_0)-t_0|+2X(\vartheta-1)I\\
&\qquad\;-(X-1)(\vartheta-1)(t_i'-t_0),(\vartheta-1)I\},
\end{align*}
as claimed.

Finally, consider time $t\in [t_i',\min\{t_{i+1}',t_1\}]$ for $i\in \mathbb{N}$.
Denote by $R_v'(t)$ the clock without applying the correction computed at time $t_i'$.
Recall that $R_v(t)$ is between $R_v'(t)$ and $R_v'(t)-(R_v(t_i')-t_i')$ during $[t_i',t_{i+1}']$.
Thus,
\begin{align*}
&\,|R_v(t)-t|\\
\le&\, \max\{|R_v'(t)-t|,|R_v'(t)-(R_v(t_i')-t_i')-t|\}\\
=&\, \max\{|R_v(t_i')+H_v(t)-H_v(t_i')-t|,\\
&\qquad\;|t_i'+H_v(t)-H_v(t_i')-t|\}\\
\le&\, |R_v(t_i')-t_i'|+|H_v(t)-H_v(t_i')-(t-t_i')|\\
\le&\, |R_v(t_i')-t_i'|+(\vartheta-1)(t-t_i')\\
\le&\, |R_v(t_i')-t_i'|+(\vartheta-1)I\\
\le&\, \max\{|R_v(t_0)-t_0|+2X(\vartheta-1)I\\
&\qquad\;-(X-1)(\vartheta-1)(t_i'-t_0),(\vartheta-1)I\}+(\vartheta-1)I\\
\le&\, \max\{|R_v(t_0)-t_0|+3X(\vartheta-1)I\\
&\qquad\;-(X-1)(\vartheta-1)(t-t_0),2(\vartheta-1)I\}.\qedhere
\end{align*}
\end{proof}
This shows that $R_v$ will converge to UTC at an amortized rate of $X-1$ when the reference clock is correct.
In particular, if the reference clock behaves nicely right from the start, we can make $|R_v(t)-t|$ arbitrarily small by choosing $I$ sufficiently small.
\greybox{
\begin{corollary}\label{cor:Roffset}
Suppose that at all times $t$ calling \textsl{ReferenceClock.Time()} at node $v$ returns $t$.
Then at all times $t$, it holds that
\begin{align*}
|R_v(t)-t|\le 2(\vartheta-1)I.
\end{align*}
\end{corollary}
}
\begin{proof}
Observe that $R_v$ is initialized to the correct time.
We now reason analogously to the proof of \Cref{lem:Roffset}, with the difference that $t_0'=t_0$, as the first loop iteration starts on initialization.
Accordingly, $R_v(t_0')-t_0'=R_v(t_0)-t_0=0$, and the resulting bound is $2(\vartheta-1)I$.  
\end{proof}
For the purpose of our analysis, denote for $v\in V_g$ and
$r\in \mathbb{N}_{>0}$ by $p_{v,r}$ the time when $v$ starts the loop for the $r$-th time. Denoting by $\vec{p}_r$ the $|V_g|$-dimensional vector whose entry for index $v\in V_g$ is $p_{v,r}$, define
\begin{equation*}
\|\vec{p}_r\|:=\max_{x\in V_g}\{p_{x,r}\} - \min_{x\in V_g}\{p_{x,r}\}.
\end{equation*}
With this terminology, our goal is to bound $\SSS:=\sup_{r\in \mathbb{N}}\{\|p_r\|\}$.

%###
\subsection{Global Synchronization}
%###
We are going to argue that \cref{alg:global_sync} can be interpreted as a variant of the classic fault-tolerant synchronization algorithm by Lynch and Welch.
Our implementation deviates from the original algorithm in a number of ways.
In part, this can be viewed as additional errors in measurements, and we can use the correctness proof of the algorithm in a black box fashion by increasing the parameters bounding the clock drift and the maximum error of phase offset measurements, i.e., $\vartheta$ and $\delta$.

Formally, abbreviate $\varrho:=(\vartheta X+1)(\vartheta-1)$ and note that by \Cref{lem:Rdrift}, there are functions $\tilde{R}_v\colon \mathbb{R}_{\ge 0}\to \mathbb{R}_{\ge 0}$, $v\in V_g$, such that for all $t'\ge t$
\begin{equation*}
|\tilde{R}_v(t')-\tilde{R}_v(t)|\le \varrho(t'-t)
\end{equation*}
and $|\tilde{R}_v(t)-R_v(t)|\le X(\vartheta-1)I$.
We will analyze the algorithm under the assumption that it runs with hardware clocks $\tilde{R}_v$.
Since the algorithm actually uses the clocks $R_v$, we account for the additive error by setting $\tilde{u}:=u+2X(\vartheta-1)I$ and $\tilde{d}:=d+X(\vartheta-1)I$ and assuming that messages are received between $\tilde{d}-\tilde{u}$ and $\tilde{d}$ time after being sent, i.e., re-interpreting this offset in the time readings as offset \emph{when} messages are sent and received.
Showing a skew of $\tilde{\SSS}$ in this setting then implies that $\SSS\le \tilde{\SSS}+2X(\vartheta-1)I$.
We remark that these additional $O((\vartheta-1)I)$ terms are marginal for the range or parameters we are interested in.

The remaining change is that we modify how clock adjustments are computed in two ways:
\begin{enumerate}
  \item If the computed clock adjustment is in the range $(-2Y(\vartheta-1)J,\delta+2Y(\vartheta-1)J)$, the clock is not modified at all.
  \item Otherwise, a clock adjustment of $Y(\vartheta-1)J$ is applied in the direction of the computed value.
\end{enumerate}
The purpose of these changes is to avoid having measurement errors introduce ``unnecessary'' clock drift into the system and, in case the reference clocks are well-behaved, not to interfere with the clocks at all.
We will show that this comes at the expense of a roughly factor $2$ increase in the worst-case skew.

The above change affects the analysis of the core algorithm, in that we need to show that the modified algorithm is able to maintain a small skew.
To this end, we will discuss here how to adapt the analysis of the algorithm provided in the lecture notes by Lenzen~\cite{lenzen2021lecture} to this setting.

Throughout this section, we make the following assumptions:
\begin{itemize}
  \item All $v\in V_g$ start their first loop iteration within $\tilde{\SSS}$ of each other with respect to the clocks $\tilde{R}_v$, i.e., $\|\vec{p}_1\|\le \tilde{\SSS}$. Since we assume that the system is initialized based on a time reference that is honest \emph{at this time,} this is easily achieved.\footnote{Nodes that join later need to obtain the current system time either from an external reference or based on offset measurements using the network. Note, however, that the latter option must be implemented in a way that does not require pre-existing synchronization of the joining node and might result in slow convergence of the joining node to UTC.}
  \item If $\|p_r\|\le \tilde{\SSS}$, the $r$-th calls to PathAwareNTPOffset for pairs $v,w\in V_g$ are executed correctly and measure phase offsets with error $\delta$, where both $\delta$ and the time required to complete the function calls depend only on $\tilde{u}$, $\tilde{d}$, $\tilde{\vartheta}$, and $\tilde{\SSS}$.
  \item At each node $v\in V_g$, the measurements are completed before local time $p_{v,r}+J-Y(\vartheta-1)J$. Observe that due to the previous assumption, this can always be achieved by choosing $J$ large enough.
  \item $X>1$ and $\varrho<1/2$. Since we want $X$ to not be too large (we choose $X=1.25$), this condition is easily satisfied; typically $\vartheta-1<10^{-4}$, which would permit values of $X$ exceeding $10^3$.
  \item $Y(\vartheta-1)\ge 2\varrho/(1-2\varrho)$, which is equivalent to
  \begin{equation}\label{eq:Y}
  2\varrho(1+Y(\vartheta-1))\le Y(\vartheta-1).
  \end{equation}
  For the relevant parameter range, this means that $Y$ must be at least about $2X$.
  \item $J\gg XI/(X-1)$.
\end{itemize}

If we can show for round $r\in \mathbb{N}_{>0}$ that $\|\vec{p}_r\|\le \tilde{\SSS}$, together with the above assumptions it follows that round $r$ is executed correctly in the sense of~\cite{lenzen2021lecture}.
In particular, denoting by $\Delta_{v,r}$ the value of \textsl{goff} node $v\in V_g$ computes in its $r$-th loop iteration, reasoning analogously as in the reference shows the following.
\greyborder{
\begin{corollary}[implicit in~\cite{lenzen2021lecture}]\label{cor:agreement}
Assume that $\|\vec{p}_r\|\le \tilde{\SSS}$. Then, for all $v\in V_g$,
\begin{equation*}
\min_{x\in V_g}\{p_{x,r}\}\le p_{v,r}+\Delta_{v,r}\le \max_{x\in V_g}\{p_{x,r}\}+\delta.
\end{equation*}
Moreover, for all $v,w\in V_g$,
\begin{equation*}
p_{v,r}+\Delta_{v,r}-(p_{w,r}+\Delta_{w,r})\le \frac{\|\vec{p}_r\|}{2}+\delta. 
\end{equation*}
\end{corollary}
}
Denote by $\tilde{\Delta}_{v,r}$ the clock adjustment node $v\in V_g$ actually applies, i.e.,
\begin{equation*}
\tilde{\Delta}_{v,r}:=\begin{cases}
Y(\vartheta-1)J & \mbox{if }\Delta_{v,r}\ge \delta+2Y(\vartheta-1)J\\
-Y(\vartheta-1)J & \mbox{if }\Delta_{v,r}\le -2Y(\vartheta-1)J\\
0 & \mbox{else.}
\end{cases}
\end{equation*}

We observe first that clock corrections serve to decrease skews.
\greyborder{
\begin{lemma}\label{lem:adjusts}
Suppose that node $v$ adjusts its clock in iteration $r$ of the modified algorithm.
Then
\begin{multline*}
\min_{w\in V_g}\{p_{w,r}\}+Y(\vartheta-1)J\le p_{v,r}+\tilde{\Delta}_{v,r}\le \\
\max_{w\in V_g}\{p_{w,r}\}-Y(\vartheta-1)J.
\end{multline*}
\end{lemma}
}
\begin{proof}
Assume first that $\tilde{\Delta}_{v,r}=Y(\vartheta-1)J$, i.e., $\Delta_{v,r}\ge \tilde{\delta}+2Y(\vartheta-1)J$.
Then
\begin{multline*}
p_{v,r}+\tilde{\Delta}_{v,r}=p_{v,r}+Y(\vartheta-1)J\ge \\
\min_{x\in V_g}\{p_{x,r}\}+Y(\vartheta-1)J.
\end{multline*}
and, by \Cref{cor:agreement},
\begin{multline*}
p_{v,r}+\tilde{\Delta}_{v,r}\le p_{v,r}+\Delta_{v,r}-Y(\vartheta-1)J-\tilde{\delta}\le \\
\max_{x\in V_g}\{p_{x,r}\}-Y(\vartheta-1)J.
\end{multline*}

The other case is that $\tilde{\Delta}_{v,r}=-Y(\vartheta-1)J$, i.e., $\Delta_{v,r}\le -2Y(\vartheta-1)J$.
Thus,
\begin{multline*}
p_{v,r}+\tilde{\Delta}_{v,r}=p_{v,r}-Y(\vartheta-1)J\le \\
\max_{x\in V_g}\{p_{x,r}\}-Y(\vartheta-1)J
\end{multline*}
and, by \Cref{cor:agreement},
\begin{multline*}
p_{v,r}+\tilde{\Delta}_{v,r}\ge p_{v,r}+\Delta_{v,r}+Y(\vartheta-1)J \ge \\
\min_{x\in V_g}\{p_{x,r}\}+Y(\vartheta-1)J.\qedhere
\end{multline*}
\end{proof}
Next, we show that large skews enforce clock corrections.
\greyborder{
\begin{lemma}\label{lem:skew}
For any $v,w\in V_g$, it holds that
\begin{align*}
&\,p_{v,r}+\tilde{\Delta}_{v,r}-(p_{w,r}+\tilde{\Delta}_{w,r})\\
\le\,& \max\{\|\vec{p}_r\|-Y(\vartheta-1)J,4\delta+10Y(\vartheta-1)J\}.
\end{align*}
\end{lemma}
}
\begin{proof}
If for any $v\in V_g$ it holds that $\tilde{\Delta}_{v,r}= 0$, we trivially have that
\begin{equation*}
\min_{x\in V_g}\{p_{x,r}\}\le p_{v,r}+\tilde{\Delta}_{v,r}=p_{v,r}\le \max_{x\in V_g}\{p_{x,r}\}.
\end{equation*}
If $\tilde{\Delta}_{v,r}\neq 0$, then \Cref{lem:adjusts} shows that
%\begin{equation*}
\begin{multline*}
\min_{x\in V_g}\{p_{x,r}\}-Y(\vartheta-1)J\le p_{v,r}+\tilde{\Delta}_{v,r}\le\\
\max_{x\in V_g}\{p_{x,r}\}+Y(\vartheta-1)J.
\end{multline*}
%\end{equation*}
Thus, if for $v,w\in V_g$ we have that $\tilde{\Delta}_{v,r}\neq 0$ or $\tilde{\Delta}_{w,r}\neq 0$, then
\begin{align*}
&\,p_{v,r}+\tilde{\Delta}_{v,r}-(p_{w,r}+\tilde{\Delta}_{w,r})\\
\le&\, \max_{x\in V_g}\{p_{x,r}\}-\min_{x\in V_g}\{p_{x,r}\}-Y(\vartheta-1)J\\
=&\,\|\vec{p}_r\|-Y(\vartheta-1)J.
\end{align*}

Hence, assume that $\tilde{\Delta}_{v,r}=\tilde{\Delta}_{w,r}=0$.
Thus,
\begin{equation*}
p_{v,r}+\tilde{\Delta}_{v,r}-(p_{w,r}+\tilde{\Delta}_{w,r})=p_{v,r}-p_{w,r}\le \|\vec{p}_r\|,
\end{equation*}
and the claim is immediate if $\|\vec{p}_r\|\le 4\delta+10Y(\vartheta-1)$.

Therefore, assume also that $\|\vec{p}_r\|>4\delta+10Y(\vartheta-1)$.
Recall that $\tilde{\Delta}_{v,r}=\tilde{\Delta}_{w,r}=0$ entails that $\Delta_{v,r}<\tilde{\delta}+2Y(\vartheta-1)J$ and $\Delta_{w,r}>-2Y(\vartheta-1)J$.
Using \Cref{cor:agreement}, we conclude that
\begin{align*}
&\,p_{v,r}+\tilde{\Delta}_{v,r}-(p_{w,r}+\tilde{\Delta}_{w,r})\\
=&\,p_{v,r}-p_{w,r}\\
<&\, p_{v,r}+\Delta_{v,r}-(p_{w,r}+\Delta_{w,r})+\delta+4Y(\vartheta-1)J\\
\le &\, \frac{\|\vec{p}_r\|}{2}+2\delta+4Y(\vartheta-1)J\\
\le &\, \|\vec{p}_r\|-Y(\vartheta-1)J.\qedhere
\end{align*}
\end{proof}
We can now bound the skew the algorithm guarantees under worst-case conditions.
\greybox{
\begin{theorem}\label{thm:worst_case_bound}
Executing \cref{alg:global_sync} alongside \cref{alg:local_sync} guarantees
\begin{equation*}
\SSS\le 4\delta+11Y(\vartheta-1)J+2X(\vartheta-1)I.
\end{equation*}
\end{theorem}
}
\begin{proof}
It is sufficient to bound $\tilde{\SSS}\le 4\delta+11Y(\vartheta-1)J$, since then $\SSS$ is bounded by
\begin{align*}
&\,\sup_{\substack{r\in \mathbb{N}_{>0}\\ v,w\in V_g}}\{\tilde{\SSS}+|R_v(p_{v,r})-\tilde{R}_v(p_{v,r})|+|R_w(p_{w,r})-\tilde{R}_w(p_{w,r})|\}\\
\le &\,\sup_{\substack{r\in \mathbb{N}_{>0}\\ v,w\in V_g}}\{\tilde{\SSS}+2X(\vartheta-1)I\}=\tilde{\SSS}+2X(\vartheta-1)I.
\end{align*}

We show this inductively, by proving that for $r\in \mathbb{N}_{>0}$, it holds that $\max_{v,w\in V_g}\{|p_{v,r}-p_{w,r}|\}=\|\vec{p}_r\|\le \tilde{\SSS}$. 
The base case is covered by our assumption that $\|\vec{p}_1\|\le \tilde{\SSS}$, so it remains to perform the step from $r$ to $r+1$.
To this end, consider $v,w\in V_g$ such that $p_{v,r+1}=\max_{x\in V_g}\{p_{x,r+1}\}$ and $p_{w,r+1}=\min_{x\in V_g}\{p_{x,r+1}\}$.
We bound
\begin{align*}
&\,p_{v,r+1}-p_{w,r+1}\\
=&\,\tilde{R}_v^{-1}(\tilde{R}_v(p_{v,r+1}))-\tilde{R}_w^{-1}(\tilde{R}_w(p_{w,r+1}))\\
=&\,\tilde{R}_v^{-1}(\tilde{R}_v(p_{v,r})+\tilde{\Delta}_{v,r}+J)\\
&\qquad\;-\tilde{R}_w^{-1}(\tilde{R}_w(p_{w,r+1})+\tilde{\Delta}_{w,r}+J)\\
\le&\, p_{v,r}+(1+\varrho)(\tilde{\Delta}_{v,r}+J)-(p_{w,r}+(1-\varrho)(\tilde{\Delta}_{w,r}+J))\\
\le&\, p_{v,r}+\tilde{\Delta}_{v,r}-(p_{w,r}+\tilde{\Delta}_{w,r})+2\varrho(\max\{\tilde{\Delta}_{v,r},\tilde{\Delta}_{w,r}\}+J)\\
\le&\, p_{v,r}+\tilde{\Delta}_{v,r}-(p_{w,r}+\tilde{\Delta}_{w,r})+2\varrho(Y(\vartheta-1)J+J)\\
\le&\, p_{v,r}+\tilde{\Delta}_{v,r}-(p_{w,r}+\tilde{\Delta}_{w,r})+Y(\vartheta-1)J,
\end{align*}
where the last step uses \eqref{eq:Y}.
Using \Cref{lem:skew}, this is bounded from above by
\begin{align*}
&\,\max\{\|\vec{p}_r\|-Y(\vartheta-1)J,4\tilde{\delta}+10Y(\vartheta-1)J\}\\
&\qquad\;+Y(\vartheta-1)J\\
= &\, \max\{\|\vec{p}_r\|,4\tilde{\delta}+11Y(\vartheta-1)J\}\\
\le &\, 4\tilde{\delta}+11Y(\vartheta-1)J.\qedhere
\end{align*}
\end{proof}

%###
\subsubsection*{Normal Operation}
%###
In order to show that the system sticks to UTC if the reference clocks accurately reproduce UTC, we first prove that \cref{alg:global_sync} does not apply any clock correction if skews are small enough.
\greyborder{
\begin{lemma}\label{lem:smallskew}
If $\|\vec{p}_r\|< 2Y(\vartheta-1)J$, then $\tilde{\Delta}_{v,r}=0$ for all $v\in V_g$.
\end{lemma}
}
\begin{proof}
Since trivially it holds that $\min_{x\in V_g}\{p_{x,r}\}\le p_{v,r}\le \max_{x\in V_g}\{p_{x,r}\}$, \Cref{cor:agreement} shows that
\begin{equation*}
\min_{x\in V_g}\{p_{x,r}\}\le p_{v,r}+\Delta_{v,r}\le \max_{x\in V_g}\{p_{x,r}\}+\Delta_{v,r}
\end{equation*}
and that
\begin{equation*}
\min_{x\in V_g}\{p_{x,r}\}+\Delta_{v,r}\le p_{v,r}+\Delta_{v,r}\le \max_{x\in V_g}\{p_{x,r}\}+\delta.
\end{equation*}
Rearranging these inequalities yields that
\begin{equation*}
-2Y(\vartheta-1)J<-\|\vec{p}_r\|=\min_{x\in V_g}\{p_{x,r}\}-\max_{x\in V_g}\{p_{x,r}\}\le \Delta_{v,r}.
\end{equation*}
and
\begin{multline*}
\Delta_{v,r}\le \max_{x\in V_g}\{p_{x,r}\}-\min_{x\in V_g}\{p_{x,r}\}+\delta= \\
\|\vec{p}_r\|+\delta<\delta+2Y(\vartheta-1)J.
\end{multline*}
Accordingly, $\tilde{\Delta}_{v,r}=0$, as claimed.
\end{proof}
It follows that if the local clock tracks UTC closely enough, no clock corrections are applied by \cref{alg:global_sync} and the skew is the one the local clocks exhibit.
\greybox{
\begin{theorem}\label{thm:good_reference}
Suppose that for some $R\in \mathbb{R}_{\ge 0}$ each $v\in V_g$ starts \cref{alg:global_sync} at time $R_v^{-1}(R)$, i.e., at local time $R$.
If at all $v\in V_g$ and times $t\in \mathbb{R}_{\ge 0}$ it holds that calling \textsl{ReferenceClock.Time()} returns $t$, then
\begin{equation*}
\SSS \le 4(\vartheta-1)I.
\end{equation*}
\end{theorem}
}
\begin{proof}
We first show that $\tilde{\Delta}_{v,r}=0$ for all $v\in V_g$ and $r\in \mathbb{N}_{>0}$.
By \Cref{cor:Roffset}, we have that $|R_v(t)-t|\le 2(\vartheta-1)I$ for all $v\in V_g$ and $t\in \mathbb{R}_{\ge 0}$.
We may assume that also $|\tilde{R}_v(t)-t|\le (\vartheta-1)I$, since this is compatible with our requirements on $\tilde{R}_v$---$X>1$ and $|\tilde{R}_v(t)-R_v(t)|$ is to be bounded by $X(\vartheta-1)I$---and our analysis applies to any suitable functions $\tilde{R}_v$.
Note that this results in $\|\vec{p}_1\|\le 2(\vartheta-1)I<2(\vartheta-1)Y$, since $p_{v,1}$, $v\in V_g$, are defined with respect to the functions $\tilde{R}_v$, and these all agree.
By \Cref{lem:smallskew}, this entails that $\tilde{\Delta}_{v,1}=0$ for all $v\in V_g$.
Straightforward induction over $r\in \mathbb{N}$ now shows that $\|\vec{p}_r\|\le 2(\vartheta-1)I$ and $\tilde{\Delta}_{v,r}=0$ for all $v\in V_g$: we just anchored the induction at $r=1$; the step consists of observing that $\|\vec{p}_{r+1}\|\le 2(\vartheta-1)I$ is immediate from the above constraint on $\tilde{R}_v$, the fact that no clock adjustment have been made so far, and applying \Cref{lem:smallskew} to conclude that $\tilde{\Delta}_{v,r+1}=0$ for all $v\in V_g$.

As we established that $\tilde{\Delta}_{v,r}=0$ for all $v\in V_g$ and $r\in \mathbb{N}_{>0}$, the $r$-th loop iteration is started at local time $R+(r-1)J$ at each $v\in V_g$.
By \Cref{cor:Roffset}, we have that the time $t_{v,r}$ with the property that $R_v(t_{v,r})=R+(j-1)J$ satisfies
\begin{equation*}
|R_v(t_{v,r})-t_{v,r}|\le 2(\vartheta-1)I.
\end{equation*}
Therefore, for any $v,w\in V_g$
\begin{align*}
|t_{v,r}-t_{w,r}|&\le |t_{v,r}-R+(r-1)J|+|R+(r-1)J-t_{w,r}|\\
&\le |t_{v,r}-R_v(t_{v,r})|+|R_v(t_{w,r})-t_{w,r}|\\
&\le 4(\vartheta-1)I,
\end{align*}
which implies the claim of the theorem.
\end{proof}
Note that there is a large slack of $2Y(\vartheta-1)J$ in the skew that is tolerated before any clock corrections are made.
Hence, any small additional deviation of $R_v(t)$ from the ``true'' UTC time caused by an imperfect reference will not change the behavior of the algorithm. 

%###
\subsubsection*{Bounded Impact of and Recovery from Reference Failure}
%###
Recall that by \Cref{lem:Rdrift}, \cref{alg:local_sync} ensures that the untrusted reference clock cannot increase the (amortized) drift of the local clock relative to real time by more than factor $\vartheta X+1$.
As shown in \Cref{lem:adjusts}, \emph{any} clock adjustment in the $r$-th iteration ensures that the adjusting node's next pulse lies within the envelope of pulse times given by $\vec{p}_r$ shifted by $J$.
Thus, a malign reference cannot increase skews with respect to UTC faster than (amortized) rate $\varrho$.
\greyborder{
\begin{lemma}\label{lem:reference_fault}
Suppose that for some $R\in \mathbb{R}_{\ge 0}$ each $v\in V_g$ starts \cref{alg:global_sync} at time $R_v^{-1}(R)$, i.e., at local time $R$.
Moreover, assume that all nodes $v\in V_g$ satisfy that $\tilde{R}_v(t)=t$ at times $t\le t_0$, i.e., up to ongoing clock corrections by \cref{alg:local_sync}, $R_v(t)$ matches UTC.
Then, for all $r\in \mathbb{N}_{>0}$ and $v\in V_g$,
\begin{equation*}
|p_{v,r}-R+(r-1)J|\le \varrho \cdot \max\{R+(r-1)J-t_0,0\}.
\end{equation*}
\end{lemma}
}
\begin{proof}
If $\tilde{\Delta}_{v,r}=0$ for some $v\in V_g$ and $r\in \mathbb{N}_{>0}$, then trivially
\begin{equation*}
\min_{x\in V_g}\{p_{x,r}\}\le p_{v,r}+\tilde{\Delta}_{v,r}\le \max_{x\in V_g}\{p_{x,r}\};
\end{equation*}
for $\tilde{\Delta}_{v,r}\neq 0$, the same inequalities follow from \Cref{lem:adjusts}.
Therefore,
\begin{multline*}
\max_{x\in V_g}\{p_{v,r+1}\}\le \\
\max_{x\in V_g}\{p_{x,r}+|\tilde{R}_v(p_{x,r+1})-\tilde{R}_v(p_{x,r})-J|+J\}.
\end{multline*}
and
\begin{multline*}
\min_{x\in V_g}\{p_{v,r+1}\}\ge \\
\min_{x\in V_g}\{p_{x,r}-|\tilde{R}_v(p_{x,r+1})-\tilde{R}_v(p_{x,r})-J|+J\}.
\end{multline*}
Since $\tilde{R}_x(t)=t$ for $t\le t_0$ and for any $t'\ge t$ we have that $|\tilde{R}_v(t')-\tilde{R}_v(t)|\le \varrho(t'-t)$, the claim of the lemma follows by induction on $r\in \mathbb{N}_{>0}$.
\end{proof}
On the other hand, once the reference clocks are functioning correctly again, the system will converge back to UTC.
\greyborder{
\begin{lemma}\label{lem:convergence}
Suppose that for $r_0\in \mathbb{N}_{>0}$ at all nodes $v\in V_g$ calling \textsl{ReferenceClock.Time()} at time $t\ge p_{v,r}$ returns $t$.
Set $\varepsilon:=3XI/J<X-1$.
Then, for all $r_0\le r\in \mathbb{N}_{>0}$ and $v\in V_g$,
\begin{align*}
|p_{v,r}-(R+(r-&1)J)|\le \max_{x\in V_g}\{|p_{x,r_0}-(R+(r_0-1)J)|\\
&\quad-(X-1-\varepsilon)(r-r_0)J,2(\vartheta-1)I\}.
\end{align*}
\end{lemma}
}
\begin{proof}
We show the claim by induction on $r_0\le r\in \mathbb{N}_{>0}$.
Since it is trivially satisfied for the base case $r=r_0$, suppose it holds for $r_0\le r\in \mathbb{N}_{>0}$ and consider the step to $r+1$.
Let $v\in V_g$ be the node maximizing $|p_{v,r+1}-(R+rJ)|$.
We perform an exhaustive case distinction.
\begin{itemize}
  \item $\tilde{\Delta}_{v,r}\neq 0$ and $p_{v,r+1}-(R+rJ)\ge 0$. By \Cref{lem:adjusts},
  \begin{multline*}
  \min_{x\in V_g}\{p_{x,r}\}+Y(\vartheta-1)J\le p_{v,r}+\tilde{\Delta}_{v,r}\le \\
  \max_{x\in V_g}\{p_{x,r}\}-Y(\vartheta-1)J.
  \end{multline*}
  We get that
  \begin{align*}
  &\, p_{v,r+1}-(R+rJ)\\
  =&\, \tilde{R}_v^{-1}(\tilde{R}_v(p_{v,r+1}))-(R+rJ)\\
  =&\, \tilde{R}_v^{-1}(\tilde{R}_v(p_{v,r}+\tilde{\Delta}_{v,r}+J))-(R+rJ)\\
  \le &\, \tilde{R}_v^{-1}(\tilde{R}_v(p_{v,r}))+(1+\varrho)(\tilde{\Delta}_{v,r}+J)-(R+rJ)\\
  \le &\, p_{v,r}+\tilde{\Delta}_{v,r}-(R+(r-1)J)+\varrho(1+Y(\vartheta-1))J\\
  \le &\, \max_{x\in V_g}\{p_{x,r}\}-(R+(r-1)J)\\
  &\qquad\;-(Y(\vartheta-1)-\varrho(1+Y(\vartheta-1)))J\\
  < &\, \max_{x\in V_g}\{p_{x,r}\}-(R+(r-1)J)-X(\vartheta-1)J\\
  < &\, \max_{x\in V_g}\{|p_{x,r_0}-(R+(r_0-1)J)|\\
  &\qquad\;-(X-1-\varepsilon)(r+1-r_0)J,2(\vartheta-1)I\},
  \end{align*}
  where the second to last inequality holds as a result of \eqref{eq:Y} and the final step uses the induction hypothesis.
  As $p_{v,r+1}-(R+rJ)\ge 0$, the induction step thus succeeds.
  \item $\tilde{\Delta}_{v,r}\neq 0$ and $p_{v,r+1}-(R+rJ)< 0$.
  Analogously to the previous case, it follows that
  \begin{align*}
  &\, p_{v,r+1}-(R+rJ)\\
  \ge &\, p_{v,r}+\tilde{\Delta}_{v,r}-(R+(r-1)J)-\varrho(1+Y(\vartheta-1))J\\
  > &\, \min_{x\in V_g}\{p_{x,r}\}-(R+(r-1)J)-X(\vartheta-1)J\\
  > &\, \min_{x\in V_g}\{-|p_{x,r_0}-(R+(r_0-1)J)|\\
  &\qquad\;+(X-1-\varepsilon)(r+1-r_0)J,2(\vartheta-1)I\}.
  \end{align*}
  As $p_{v,r+1}-(R+rJ)< 0$, the induction step thus succeeds.
  \item $\tilde{\Delta}_{v,r}=0$.
  Denote by $L_v(t)$ the value returned by calling \textsl{LocalClock.Time()} at node $v$ at time $t$.
  Since the algorithm always sleeps for $J$ time in each loop iteration, where we assume computations to take zero time, we have that $L_v(p_{v,r+1})=R+rJ$ and $L_v(p_{v,r})=R+(r-1)J$.
  We apply \Cref{lem:Roffset} with $t_0=p_{v,r}$ and $t_1=t=p_{v,r+1}$, noting that the lemma is applicable, since during this time \cref{alg:global_sync} does not interfere with the local clock.
  Thus, the lemma shows that
  \begin{align*}
  &\,|p_{v,r+1}-(R+rJ)|\\
  =&\,|L_v(p_{v,r+1})-p_{v,r+1}|\\
  \le &\, \max\{|L_v(p_{v,r})-p_{v,r}|\\
  &\qquad\;+3X(\vartheta-1)I-(X-1)(\vartheta-1)J,2(\vartheta-1)I\}\\
  = &\,\max\{|p_{v,r}-(R+(r-1)J)|\\
  &\qquad\;-(X-1-\varepsilon)(\vartheta-1)J,2(\vartheta-1)I\}\\
  \le &\,\max_{x\in V_g}\{|p_{x,r_0}-(R+(r_0-1)J)|\\
  &\qquad\;-(X-1-\varepsilon)(\vartheta-1)(r+1-r_0)J,2(\vartheta-1)I\},
  \end{align*}
  completing the induction step also in this case.\qedhere
\end{itemize}
\end{proof}
Together, we arrive at the following result.
\greybox{
\begin{theorem}\label{thm:convergence}
Suppose that at initialization time, the reference clocks return the correct UTC time.
During periods where the reference clocks are faulty or malicious, local times may drift from UTC at rate at most $\varrho=(\vartheta X+1)(\vartheta-1)$.
During periods where the reference clocks return UTC, any resulting deviation is reduced at amortized rate $X-1-\varepsilon$, where $\varepsilon=3XI/J$.
Once the (re-)convergence process is complete, the skew is bounded by $4(\vartheta-1)I$ (until the reference clocks fail again).
\end{theorem}
}
\begin{proof}
The first two statements of the theorem paraphrase \Cref{lem:reference_fault,lem:convergence}.
Regarding the third, note that once the bound of \Cref{lem:convergence} is attained by the second term in the maximum, we have for $v,w\in V_g$ and the indices $r$ that
\begin{equation*}
|p_{v,r}-p_{w,r}|\le |p_{v,r}-t|+|p_{w,r}-t|\le 4(\vartheta-1)I.\qedhere
\end{equation*}
\end{proof}

\else
\fi

\section{Evaluation}
\label{sec:eval}
In this section, we evaluate the reliability of \name under various proportions of malicious in-network attackers. We then compare different path selection strategies and the number of paths being used in the core time synchronization. In addition, we investigate the availability of \name without GNSSes, demonstrating resilience to long-term failure of the reference clocks. At last, we analyze the additional scalability requirements for core time servers participating in \name.

\subsection{Experimental Setup}
\label{ssec:experimentalsetup}
Despite the ever-expanding commercial deployment of SCION in the past few years~\cite{cyrill2021deployment}, the global production network
is not yet sufficiently diverse for our experiments. Therefore, we conduct our experiments by simulating \name on realistic inter-domain topologies consisting of either the 2000 or the 500 highest-degree Tier-1 and Tier-2 ASes, extracted from the CAIDA AS relationships with geographic locations data set~\cite{CAIDA-Data-Geo}. To this end, we have developed an ns-3-based simulator~\cite{nsnam_ns-3_nodate} that supports the SCION control and data plane. We conduct all experiments on the topology of 2000 ASes except long-term (one year) experiments, where we conduct them on the topology of 500 ASes due to execution time constraints.

The control plane provides ASes with diverse paths to any destination AS. Since the topology is massively interconnected, the longest distance between any pair of ASes is five. Each AS stores at most 60 different paths per destination AS. In the data plane, we simulate propagation delay by calculating the great circle latency between border routers of an AS using their locations specified by the CAIDA AS data set. For the propagation delay between two adjacent routers at the same location we assume a 1 meter long optical fiber between them. Furthermore, we simulate the transmission delay by assuming 400 Gbps links between all border routers, and queuing delay by assuming each router can process 5 Gpps. We run the simulations in a compute cluster on 64 CPU cores with 32 or 120 GB of memory, depending on problem size.

\subsection{Reliability Analysis} % reliability
\label{ssec:reliability}
To evaluate the reliability of \name, we specifically consider an on-path delay attacker model among the various attack methods, because a simple authentication scheme can mitigate other attacks. Furthermore, an off-path delay attack requires complex preceding attacks (e.g., BGP hijacking) manipulating the routing infrastructure to introduce arbitrary asymmetric delays in packet exchanges, which is impossible in path-aware networking environments such as SCION. 

An on-path attacker, however, can impact the packets passing through nodes under its control. It is possible to enlarge the attack influence by strategically controlling core ASes with the highest degree of connectivity, but compromising such ASes is assumed to be very difficult in realistic scenarios. In addition, ASes in SCION can quickly switch the forwarding path over a different route as soon as they observe any suspicious behavior of the compromised ASes. Driven by this, we consider a botnet-size attacker distributed uniformly at random throughout the network.

We conduct 20 different experiments on the topology of 2000 ASes with a combination of four different attacker populations 
(i.e., 5, 10, 15, and 20\%) and five different path selection strategies for non-neighboring 
ASes (i.e., $k$ shortest, disjoint, and random paths). In all the experiments, each AS is
assigned with a uniformly and independently random drift in a range of $\pm$ \SI{27}{\mu\second} per day.
The global cutoff is \SI{1}{m\second}, \textsl{LocalClock.MaxDrift} is \SI{27}{\mu\second} per day,
and the coefficients $X$ and $Y$ are 1.25 and 2.5, respectively.
The attacker ASes introduce a random asymmetry in the range of 50 ms to 300 ms between their own border router pairs.

\begin{figure}[t]
\subfloat[5\% of attacker nodes.]{%
\centering
\includegraphics[width=0.48\linewidth]{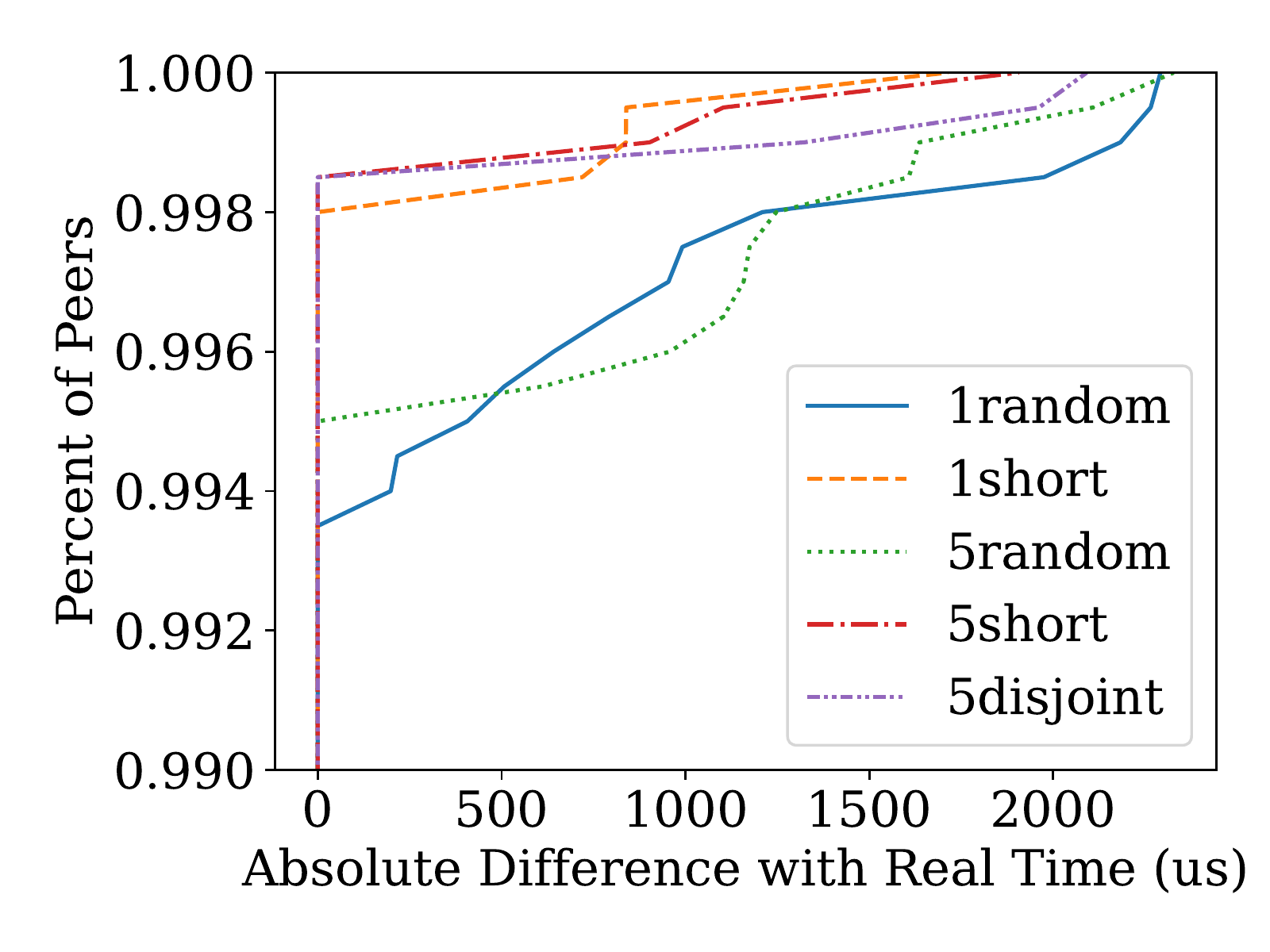}
\label{sfig:5percent}
}
\subfloat[10\% of attacker nodes.]{%
\centering
\includegraphics[width=0.48\linewidth]{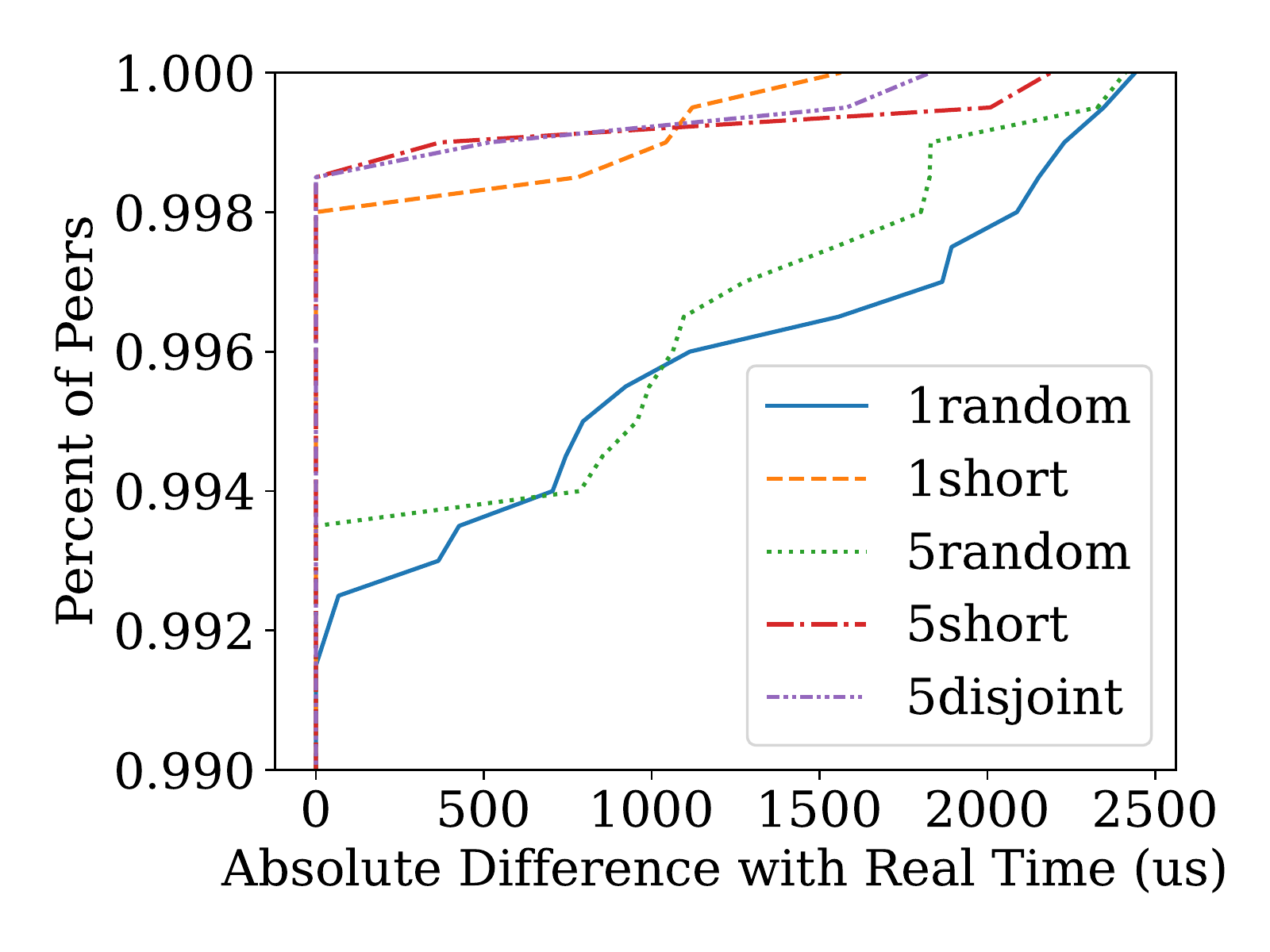}
\label{sfig:10percent}
}\newline
\subfloat[15\% of attacker nodes.]{%
\centering
\includegraphics[width=0.48\linewidth]{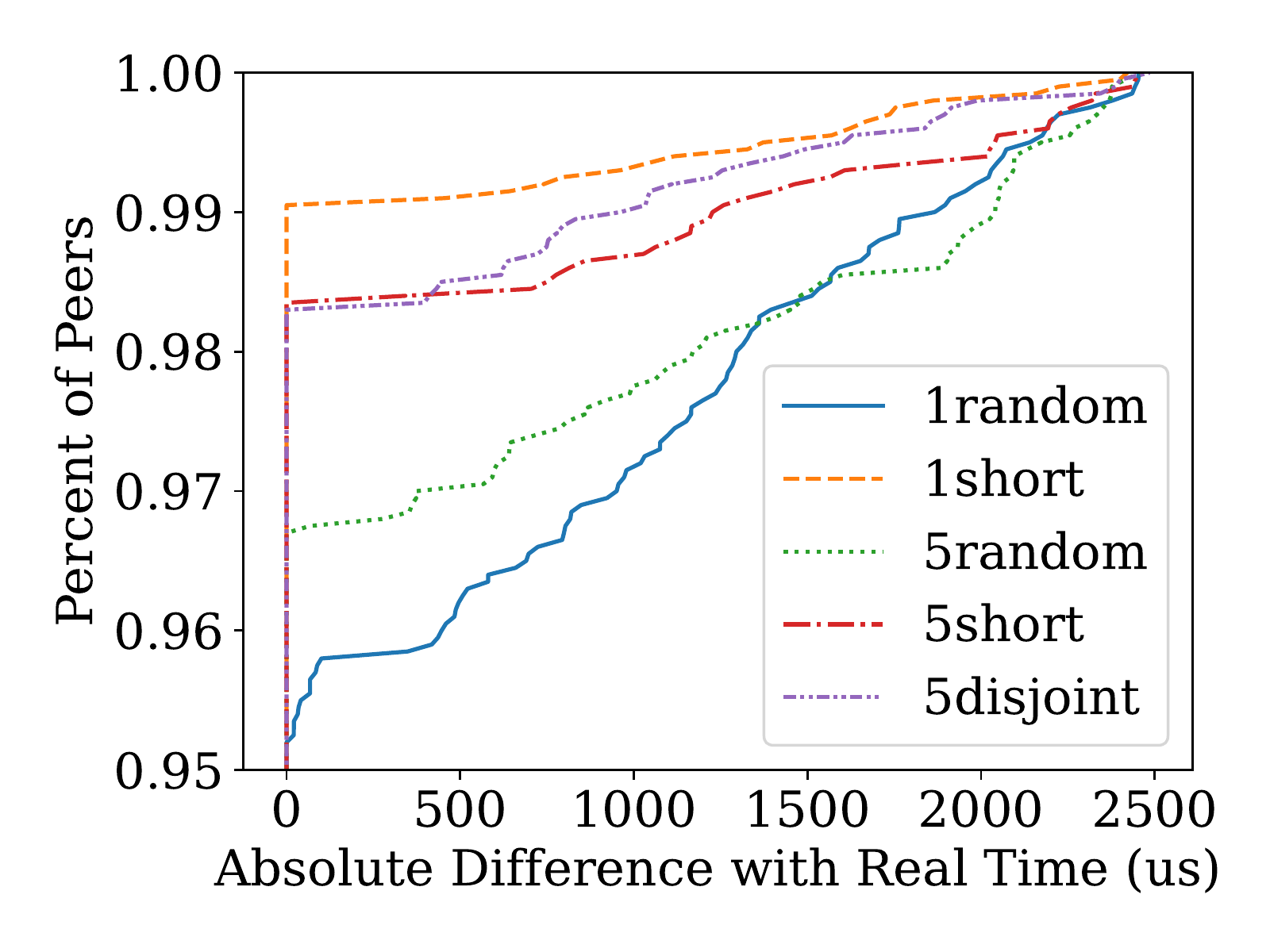}
\label{sfig:15percent}
}
\subfloat[20\% of attacker nodes.]{%
\centering
\includegraphics[width=0.48\linewidth]{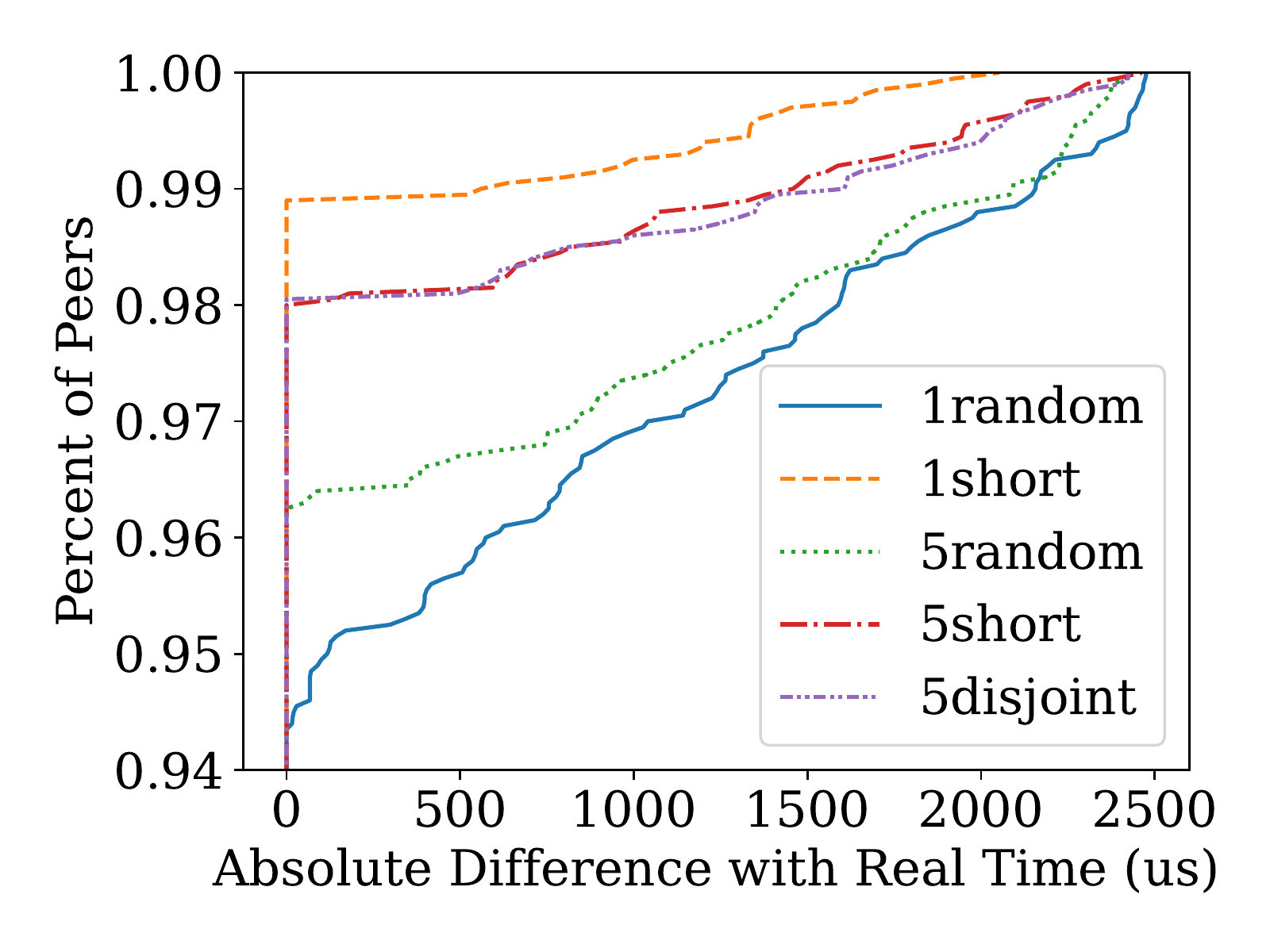}
\label{sfig:20percent}
}
\caption{CDF of local clock offsets to real-time according to different proportions of attackers in a network of 2000 ASes.}
\label{fig:reliability}
\end{figure}

\paragraph{Overall Performance} % single path vs. multipath
\cref{fig:reliability} shows cumulative distribution function (CDF) results of local clock offsets at each AS to real-time over 40 days.  
The majority of ASes achieve highly reliable global time synchronization with only a few microseconds of deviation from real-time; more precisely, even in the extreme case where 20\% of entities in the network act maliciously, over 94\% of ASes are closely synchronized with real-time. In comparison, 99.3\% of ASes are successfully synchronized with real-time when 5\% of nodes are malicious. Although some ASes exhibit desynchronization by the delay attack, thanks to \textsl{maxCorr}, they avoid substantial time-shifts, resulting in a maximum of about \SI{2500}{\mu\second} after 40 days.
An interesting observation is that multipath does not always guarantee better performance than the shortest path. However, in realistic cases where only small fractions of the network are compromised (e.g., $\leq$ 10\%), attackers can desynchronize fewer ASes when a multipath strategy is used.

\paragraph{Path Selection Analysis} % analysis on path selection strategies. 
Another interesting point to highlight is that path selection strategies can be a key factor in overall performance. Each AS uses the single shortest path towards its neighboring ASes, while it applies three different path selection strategies for non-neighboring ASes: shortest path, disjoint path, and random path. In the shortest path selection, each AS evaluates the number of hops of the  forwarding paths towards a destination AS provided by the control plane and selects $k$ shortest paths (i.e., $k = 5$ for multipath, otherwise $k = 1$). In the case of disjoint path selection, similar to the shortest path selection, each AS evaluates the set of paths provided by the control plane and selects the shortest paths first. For the second path, however, it selects the most disjoint path with respect to previously selected paths and continues until all $k$ shortest disjoint paths are determined. Finally, the random path selection performs uniform sampling from the set of paths provided by SCION.

\Cref{fig:reliability} demonstrates the benefit of multipath strategies; using multiple short or disjoint paths, fewer ASes are affected by an attacker population of up to 10\%. However, the longer tail of the curves for the multipath strategies indicate that \emph{affected ASes} could deviate more from real-time when they use multiple paths. Furthermore, although random path selection strategies (multi- or singlepath) indicate inferior performance compared to other strategies, in all attack scenarios using multiple random paths significantly reduces the number of affected ASes in comparison to using one random path. Despite the decent performance of multipath strategies when the attacker percentage is 5\% or 10\%, increasing this proportion to 15\% and 20\% deteriorates their performance relative to the (single) shortest path strategy. That is because, in a network with sparse attacker distribution, the probability of avoiding the attacker nodes through multiple completely different paths is high, whereas this probability decreases in a network with densely distributed attacker nodes.

\begin{figure}[t]
\subfloat[The shortest path]{%
\centering
\includegraphics[width=0.48\linewidth]{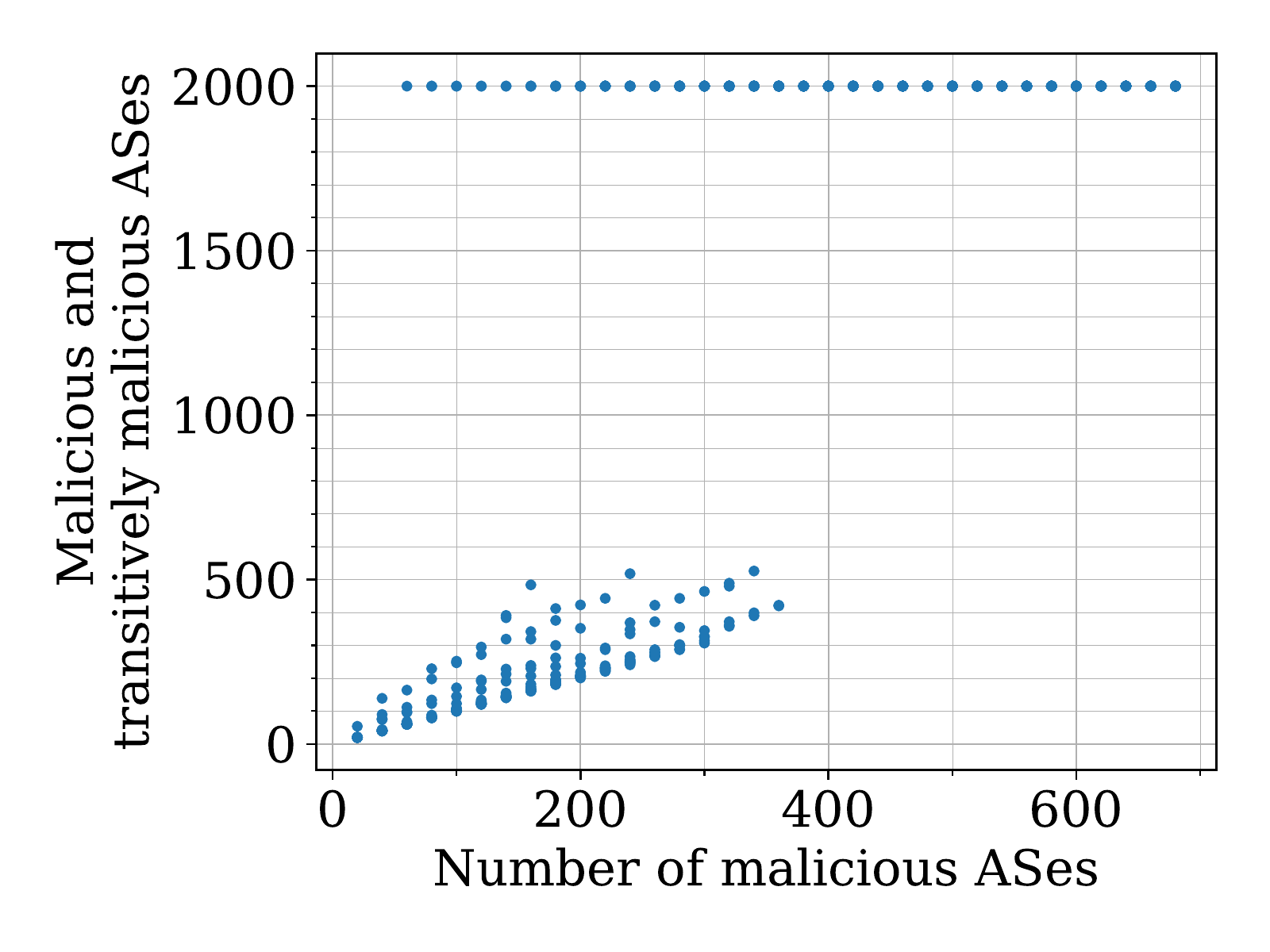}
\label{sfig:1shotest}
}
\subfloat[One random path]{%
\centering
\includegraphics[width=0.48\linewidth]{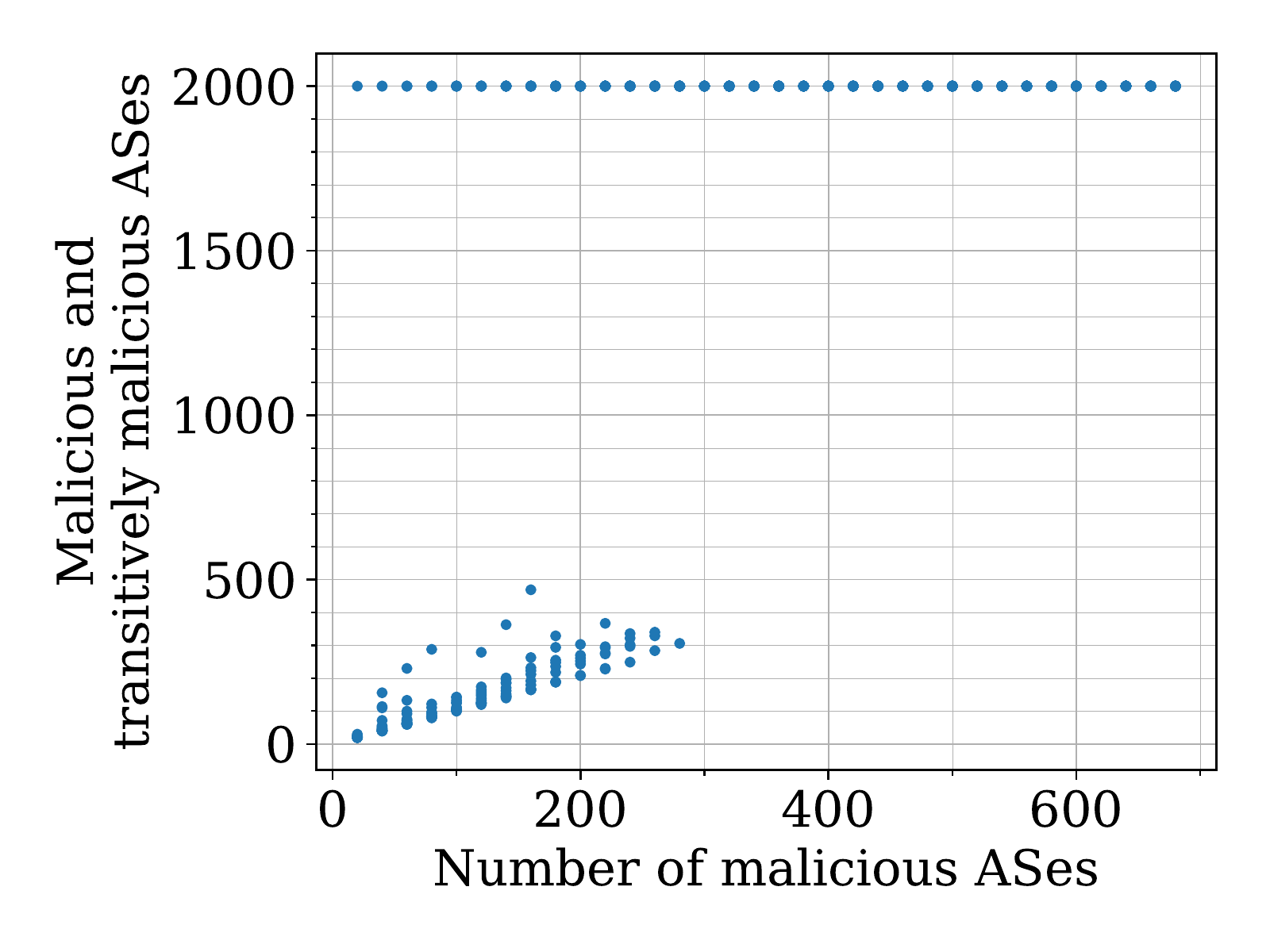}
\label{sfig:1random}
}\newline
\subfloat[Five shortest paths]{%
\centering
\includegraphics[width=0.48\linewidth]{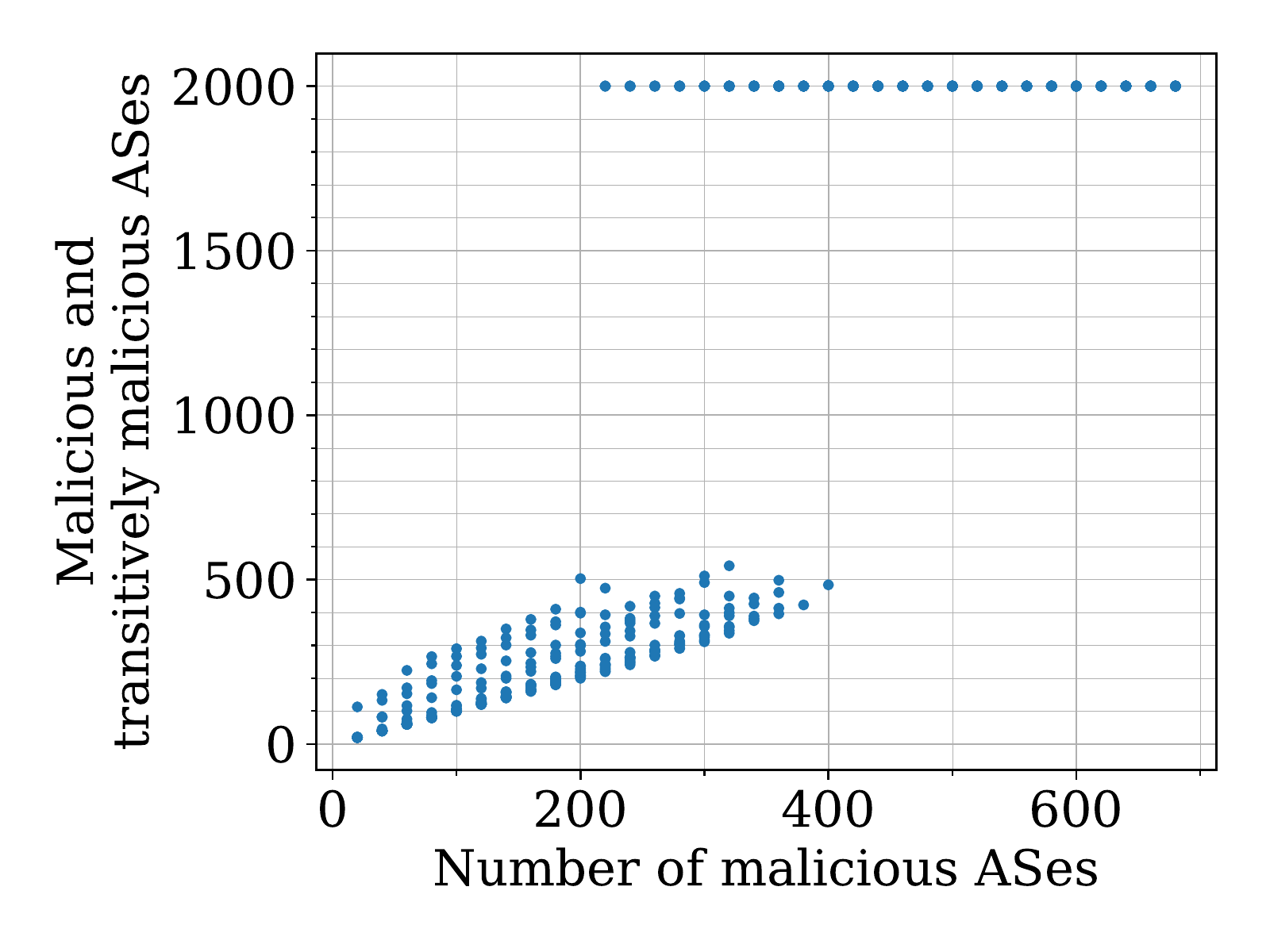}
\label{sfig:5shotest}
}
\subfloat[Five random paths]{%
\centering
\includegraphics[width=0.48\linewidth]{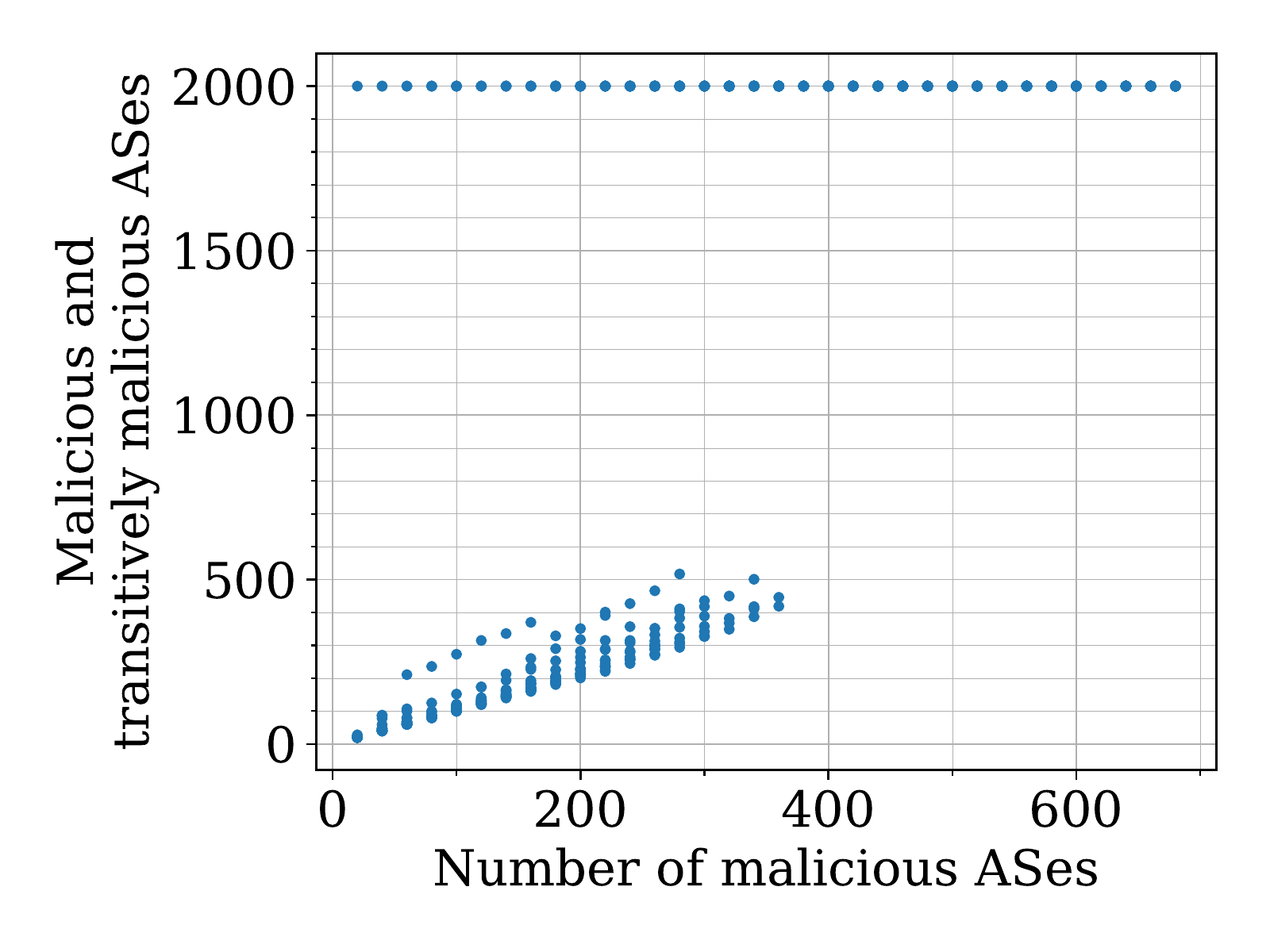}
\label{sfig:5random}
}
\caption{Number of maliciously affected ASes for different path selection strategies in a topology of 2000 ASes.}
\label{fig:path_analysis}
\end{figure}

\paragraph{Worst-Case Behavior of Compromised Nodes}
Furthermore, we conduct a reliability analysis only based on the presence of attackers on the paths selected by peers. In order to bound the effect the attacker can have on the system in the worst case, we consider an AS transitively corrupted if the attacker might effectively gain control of its output clock. Once the attacker can arbitrarily pull the clock of an AS, this might in turn affect other ASes, leading to a domino effect.

To reflect this, we repeatedly uniformly sample sets of attacker ASes, ranging from none to one third of the nodes. In each iteration, we determine which ASes might be transitively affected. An AS is affected (i.e., desynchronized) by attackers or transitively corrupted ASes if the fault-tolerant midpoint calculation of the algorithm uses one third or more corrupt input values. Each input is given by measuring the offset to another AS. A measurement is corrupt if the majority of paths used contain a node controlled by the attacker, or if the AS to which the offset is measured is itself (transitively) corrupted. \Cref{fig:path_analysis} shows the relation between the number of malicious ASes (x-axis) and the number of malicious and transitively corrupted ASes for different path selection policies.

The plots show essentially three regions: below a certain threshold only a few ASes are transitively corrupted, above a larger threshold the system globally fails, and between the two thresholds either case might apply. Using five shortest paths, up to $400$, i.e., about $20\%$ of primary corrupted nodes can be sustained reliably. Moreover, this figure indicates that using more than one path increases the number of required attackers to affect the whole network.

\begin{figure}[t]
\subfloat[No references over a year]{%
\centering
\includegraphics[width=0.48\linewidth]{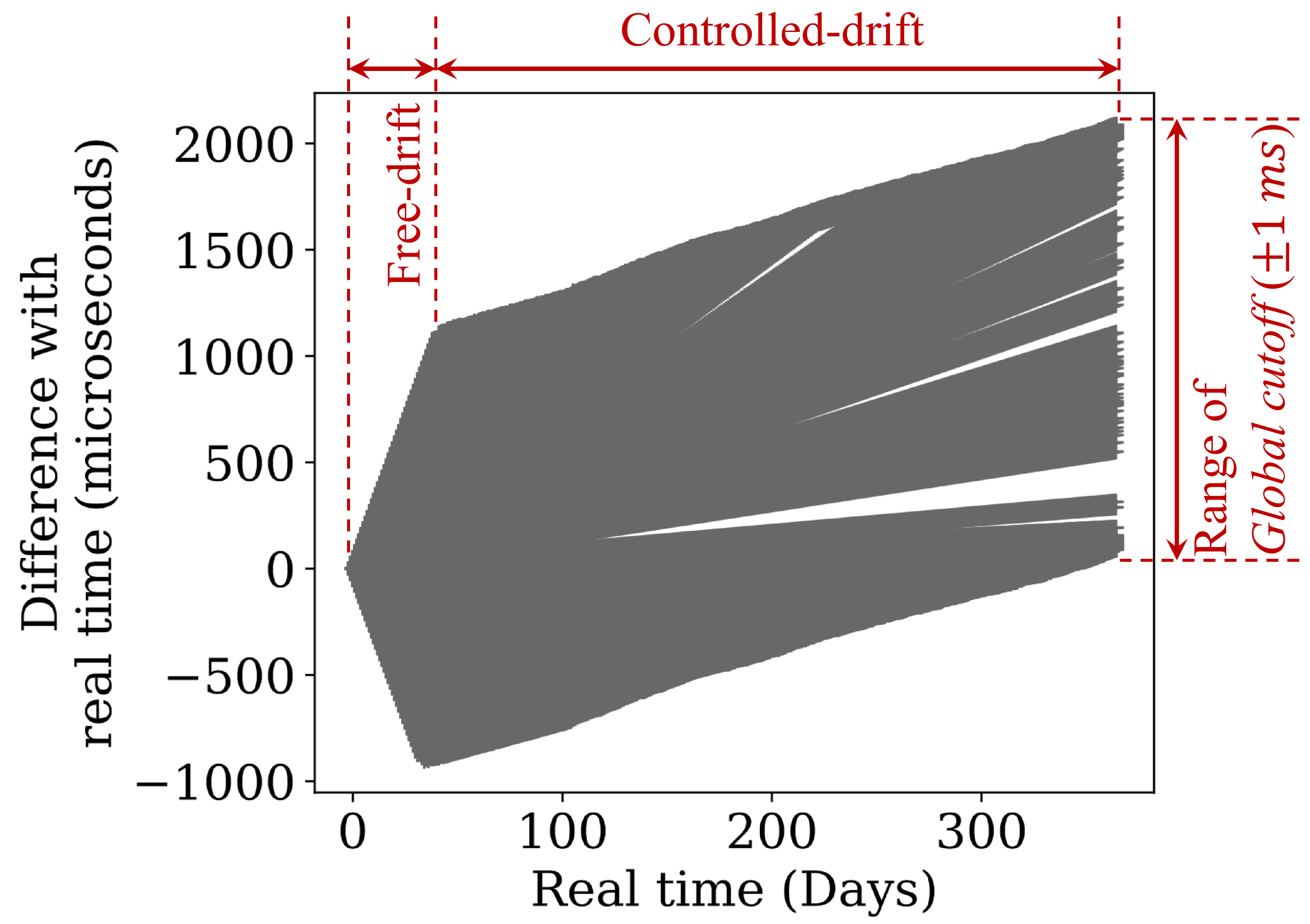}
\label{sfig:availability}
}
\subfloat[No references from day 100 to 200]{%
\centering
\includegraphics[width=0.48\linewidth]{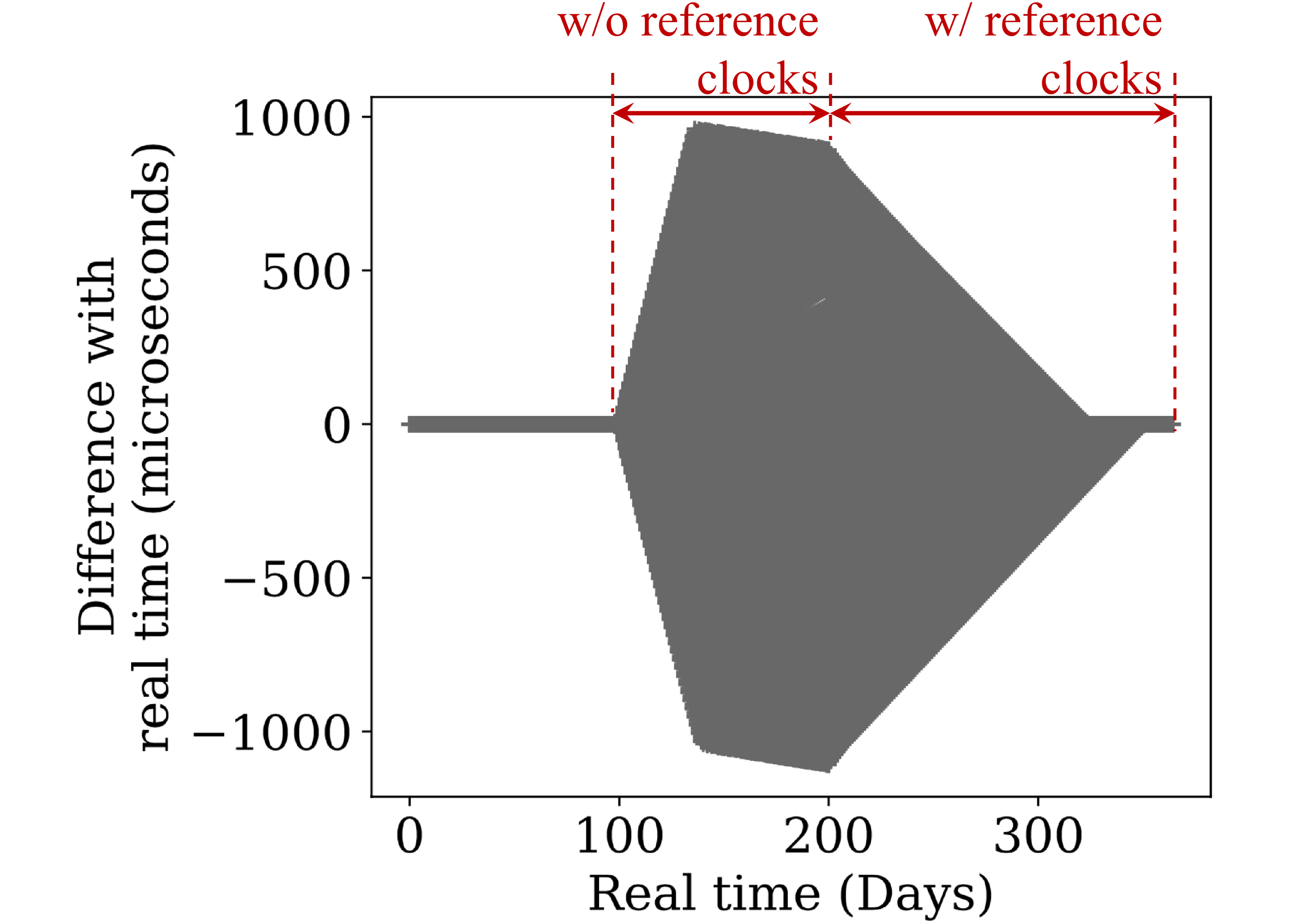}
\label{sfig:availability2}
}
\caption{\name results with and without external reference clocks in a network of 500 ASes.}
\label{fig:availability}
\end{figure}

\subsection{Availability Analysis}
\label{ssec:availability}
We now evaluate the long-term availability of \name with no external reference clocks, simulating an extended GNSS outage caused by, e.g., a solar superstorm~\cite{superstorm2021}; we assume that some underground cables may also be damaged, but thanks to the multipath infrastructure, connectivity between all ASes is assumed to be intact. In this simulation scenario, each local clock solely depends on the global clock synchronization, and no local clock adjustment with the reference clock is available. 
Our objective is thus to tie all the peers within a synchronization threshold to each other, not to real-time.
We set \textsl{LocalClock.MaxDrift} to \SI{27}{\mu\second} per day and global cutoff~$=$~\SI{1}{ms}.
\cref{sfig:availability} depicts the \name result for a year, and \cref{sfig:availability2} depicts the results for a scenario where reference clocks disappear at some point and come back online again after some time.

From the results we observe the following: 
(i) For the first month, ASes drift apart following their local drifts (free-drift period). Since each AS has a different local clock drift, their local times are freely distributed (in a radial pattern) within the global cutoff range.
(ii) ASes that reached global cutoff do not drift further apart, and as a result, all ASes remain within a range of $\pm$ \SI{1}{ms} (controlled-drift period). 
(iii) The global time inevitably biases because the global clock synchronization process is affected by the delay asymmetry in packet switching, different local clock drifts, and different synchronization initiation times.
Finally, 
ASes are eventually classified into upper-bound group (local drift > global drift), lower bound group (local drift < global drift), and centerline group (local drift = global drift).
Nevertheless, \name successfully ties together all the core ASes within \SI{2}{ms} of time drift from real-time for a year of reference clock outage. 

As shown in~\cref{sfig:availability2}, once the references are available again, the system converges back to UTC within a time span comparable to the outage. The speed of convergence is governed by the restricted corrections that are applied in the clock synchronization algorithm. Hence, the system reconverges more quickly after short outages.

\subsection{Scalability Analysis}
For the current design iteration, we assume that the set of core \ts{es} will eventually encompass about 2'000 nodes. A critical question, therefore, is whether the core \ts{es} will be able to (vertically) scale with the expected NTP traffic for the global synchronization as well as the traffic caused by intra-ISD synchronization requests. Computing a Byzantine fault-tolerant approximate agreement on clock corrections based on Lynch and Welch requires a single message exchange from each core \ts{} to all its peers. To make optimal use of the path-aware networking infrastructure, we anticipate conducting offset measurements between any pair of peers in parallel over up to 5 different paths. This amounts to about 10'000 message exchanges that every core \ts{} has to handle in each round.
Based on the formal analysis \ifnum\showanalysis=1 in~\cref{sec:formal-analysis}\else in~\cite{g-sinc-extended-2022}\fi, it will be sufficient to schedule global synchronizations with an interval on the order of hours, e.g., once every hour.

We investigate the scalability question in a simple micro-benchmark between a server machine running our implementation of a SCION-based NTP server process and a load generator machine for NTP client requests via SCION. Both machines were running the operating system Ubuntu Server 20.04 LTS on Amazon's Elastic Compute Cloud instances equipped with 64 (virtual) CPUs (64-bit ARM), 256 GiB of memory, and 25 Gbps of available network bandwidth. Repeatable tests show that our implementation is able to sustain an average load of at least 370'000 NTP requests per second in one process without packet loss. Peak server performance was measured with values of over 400'000 NTP requests per second using the network monitoring tool bwm-ng~\cite{bwm-ng}. These results indicate that our implementation is practical despite the additional SCION packet processing overhead.

\section{Related Work}
\label{sec:related}
Extensive studies have been performed on the fundamental problem of clock synchronization in distributed systems to improve accuracy, security, and reliability.

\paragraph{Accuracy}
The \emph{Precision Time Protocol (PTP)} uses hardware timestamping to eliminate network stack delays~\cite{ieee1588}.
Thanks to the extensive hardware support at switches, PTP is known to be able to achieve sub-microsecond on a LAN and even sub-nanosecond precision in a well-provisioned datacenter~\cite{moreira2009}.
Nevertheless, to guarantee the high precision, the network needs to be fully PTP-enabled; otherwise, the precision will significantly degrade~\cite{zarick2011transparent,lee2016globally}.
Both \emph{Datacenter Time Protocol (DTP)}~\cite{lee2016globally} and HUYGENS~\cite{geng2018} can achieve synchronization accuracy of a few 10s of nanoseconds but both systems are targeted only at deployments in datacenters. 

\paragraph{Security}
Although a large volume of work is carried out on clock synchronization to improve accuracy, its security has only recently gotten attention~\cite{malhotra2016attacking, rfc-8633, Jeitner2020}.

Early NTP had no security design aspect.
\emph{NTPv3} had first adopted packet authentication with a pre-shared symmetric key, which needs to be established out-of-band~\cite{rfc-1305}.
In \emph{NTPv4}, a PKI-based authentication mechanism has been introduced~\cite{rfc-5905}.
\emph{IEEE 1588}'s experimental security extension describes an HMAC-based authentication method for PTP, and later several variants (e.g., GMAC or CMAC) were implemented and tested~\cite{hirschler2011validation, onal2012security}.
Unfortunately, they have shown only limited adoption in practice due to the overhead of server-side public key operations.
\emph{Authenticated Network Time Synchronization (ANTP)} aims at large-scale deployments~\cite{dowling2016authenticated}.
It minimizes server-side cryptographic operations using symmetric cryptography for subsequent synchronization processes while enabling the servers to be stateless.
\emph{SecureTime} employs high-performance digital signature schemes to secure multicast time synchronization~\cite{annessi2017s}.
The \emph{Secure Time Synchronization (STS)} protocol offloads the authorization to a third party (i.e., Authorization Server), resolving the circular dependency between time synchronization authentication and certificate validation~\cite{mkacher2018secure}.

The \emph{NTP Pool Project} provides more than 4'500 NTP servers (as of April 2022), with which hundreds of millions of systems are able to sync, mitigating individual server failures.
Each NTP client gathers clock samples from multiple NTP servers and selects the best clock samples to update the local clock~\cite{ntppoolproject}.
Perry et al. demonstrate that injecting malicious timeservers into the NTP pool is alarmingly feasible, influencing a significantly large number of systems~\cite{perry2021devil}.
\emph{Chronos} applies an approximate agreement algorithm to a large number of NTP servers to guarantee reliable time synchronization even in case many time servers are faulty or under an attacker's control~\cite{deutsch2018chronos}.

\paragraph{Reliability}
A number of projects and publications with the specific goal of assessing and increasing the robustness of GNSS-based clock synchronization have already been discussed in~\cref{sec:intro2}. The experimental \textsl{PTP ring method} outlined in~\cite{Sherman2021} enables time servers with disrupted GNSS clocks to access a redundant source of frequency from clocks located elsewhere in the network. The underlying motivation is related to the contributions of \name despite the fact that direct synchronization via individual PTP links is unable to provide the same degree of (Byzantine) fault tolerance.

\section{Conclusion}
\label{sec:conclusion}
It is challenging to design a system with the ambition of going beyond what established time synchronization architectures provide to support critical infrastructure. This is a testimony showing how well for example NTP has been developed over many iterations into an architecture that fulfills its basic requirements to a large degree. Yet \name demonstrates that it is possible to enhance the current state of the art in clock synchronization in particular when it comes to essential qualities like global scalability without a single root of trust, mitigation of network-level attacks, and in general availability even under extreme conditions. This is achieved by building on the solid body of fault-tolerant clock synchronization research and the SCION Internet architecture providing strong resilience and security properties as an intrinsic consequence of its underlying design principles.

\section{Acknowledgements}
\label{sec:acknowledgements}
We would like to thank the anonymous reviewers for their insightful feedback and valuable suggestions. In particular, we would like to express our gratitude to this paper's shepherd, Alysson Bessani, who helped improve it substantially. Sincere thanks go also to Attila Kinali, David Kleymann, Filip Meier, and Zilin Wang for their generous contributions to the project at various stages of development. We gratefully acknowledge support from ETH Zurich, and from the Zurich Information Security and Privacy Center (ZISC). This project has received funding from the European Research Council (ERC) under the European Union's Horizon 2020 research and innovation programme (grant agreement 716562). We extend our profound thanks for this contribution.

\label{end:paper}

\bibliographystyle{IEEEtranS}
% Generated by IEEEtranS.bst, version: 1.14 (2015/08/26)

\label{end:references}
\end{document}